\documentclass[onecolumn,11pt]{article}
\usepackage{amssymb,amsmath,epsfig}

\begin{document}

\title{\bf Bouncing Cosmology for Entropy Corrected Models in Ho$\check{\text r}$ava-Lifshitz
Gravity and Fractal Universe}

\author{\bf{Tanwi Bandyopadhyay}
\thanks{tanwi.bandyopadhyay@aiim.ac.in}\\
Department of Mathematics, Adani Institute of Infrastructure
Engineering,\\ Ahmedabad-382421, Gujarat, India.\\\\
\bf{Ujjal Debnath}
\thanks{ujjaldebnath@gmail.com} \\
Department of Mathematics, Indian Institute of Engineering\\
Science and Technology, Shibpur, Howrah-711 103, India.}

\maketitle

\begin{abstract}
The modified field equations are written in logarithmic and power
law versions of entropy corrected models in Einstein's gravity in
the background of FRW universe. In one section, a brief review of
the Ho$\check{\text r}$ava-Lifshitz gravity is discussed and the
modified field equations in logarithmic and power law versions of
entropy corrected models in Ho$\check{\text r}$ava-Lifshitz
gravity are formulated. The stability analysis for these models
are performed by describing the dynamical system. In another
section, a brief review of the fractal universe is presented and
the modified field equations in logarithmic and power law versions
of entropy corrected models in fractal universe are formulated.
The stability analysis for the dynamical system for these models
in the framework of fractal universe are described. Furthermore,
the bouncing scenarios of the universe in Ho$\check{\text
r}$ava-Lifshitz gravity and fractal model for both logarithmic and
power law entropy corrected versions in $k=0,+1,-1$ separately are
analyzed. For different cases, the validity of null energy
condition (NEC) at the time of bounce is examined. Finally, the
behaviors of the physical quantities are depicted through diagrams. \\

\end{abstract}

\sloppy \tableofcontents

\section{Introduction}

In the General Relativity (GR), the initial big bang singularity
is the main problem in cosmological models \cite{Mukh}. Many
theoretical inconsistencies have been occurred in the big bang
cosmology. The consistent framework for this early time cosmology
is the inflationary paradigm \cite{Mukh1,Mukh2,Lyth,Lind}.
However, the inflation theories suffer from the initial
singularity problem at the origin $t = 0$ \cite{Bord}. To resolve
the initial singularity, the modification of GR theory
\cite{Novel} has been proposed (for example, Superstring theory,
loop quantum gravity etc). To avoid the big bang singularity, an
appealing alternative scenario is the big bounce scenario, where
the singularity is replaced by a bounce \cite{Mukh3,Odin}. Another
alternative to standard big bang cosmology (to avoid the big bang
singularity) is oscillating universe where the big bang is
replaced by cyclical evolution \cite{Tolman}. In the literatures,
several authors have been studied the bouncing cosmology
\cite{Xu,Noj,Ant,Geg,Cai0}. The bouncing cosmology has been
studied in modified gravity theories eg. $f(R)$ gravity
\cite{Carl,Barr,Paul,Ghan,Fard,Odin1,Odin2}, $f(T)$ gravity
\cite{Cai00,Asta}, loop quantum cosmology \cite{Boj,Singh,Haro},
braneworld gravity \cite{Ste1,Ste2,Piao1,Piao2}, Brans-Dicke
theory \cite{Sing1}, Gauss-Bonnet gravity \cite{Bam,Oik,Te,Es}.
Page \cite{Page} has discussed the fractal set of perpetually
bouncing universes. Rip singularity scenario and bouncing universe
in Chaplygin gas model have been studied in \cite{Sada}. Salehi
\cite{Sal} has studied the bouncing universe in presence of
extended Chaplygin gas model. Also the study on the bouncing
behavior of modified Chaplygin gas in presence of bulk viscosity
has been discussed in \cite{Chat}. Recently, we have studied
\cite{Band} the bouncing scenarios of the universe in the
framework of generalized cosmic Chaplygin gas and variable
modified Chaplygin gas. In the literature, the bouncing nature of
the universe in several kinds of dark energy models have been
investigated \cite{Col,Arab,Maka,Bram,Aly}. Motivated by the above
works, we investigate the bouncing cosmology for entropy corrected
dark energy models in the framework of Ho$\check{\text
r}$ava-Lifshitz gravity and fractal universe.\\

The organization of the paper is as follows: In section 2, we
formulate the modified field equations in logarithmic and power
law versions of entropy corrected models. In section 3, we briefly
review the Ho$\check{\text r}$ava-Lifshitz gravity. In section 4,
we write the modified field equations in logarithmic and power law
versions of entropy corrected models in Ho$\check{\text
r}$ava-Lifshitz gravity. Then we describe the dynamical system and
stability analysis for these models. In section 5, we present a
short review of the fractal universe. In section 6, we write the
modified field equations in logarithmic and power law versions of
entropy corrected models in fractal universe. Then we describe the
dynamical system and stability analysis for these models in the
framework of fractal universe. In section 7, we study the bouncing
scenarios of the universe in Ho$\check{\text r}$ava-Lifshitz
gravity and fractal model for both logarithmic and power law
entropy corrected versions. Finally, we present the conclusion in
section 8.

\section{Modified Equations for Entropy Corrected Models in Einstein's Gravity}

In the quantum field theory, the semi-classical quantum properties
of black hole can be analyzed in curved backgrounds. Here matter
is described by quantum field theory and gravity is described as a
classical background. In this context, it is well known that the
black hole can emit Hawking radiation where the temperature is
proportional to its surface gravity on the black hole horizon and
the entropy of the black hole is proportional to its horizon area.
The black hole entropy $S$ plays an important role to study the
astrophysical research, where $S=\frac{A}{4G}$. Here $A$ is the
area of the black hole horizon. In recent years, the quantum
corrections (logarithmic and power law corrections) to the
semi-classical entropy law have been introduced due to thermal
equilibrium fluctuation or quantum fluctuation.

\subsection{Modified Equations from Logarithmic Corrected Entropy}

The logarithmic correction of entropy had been incorporated due to
the thermal equilibrium or quantum fluctuations
\cite{Mann,Meis,Chatt,Mod,Sad}. The Entropy-area relation in
logarithmic correction of entropy is given by

\begin{equation}
S=\frac{A}{4G}+\alpha\ln \frac{A}{4G}+\beta \frac{4G}{A}
\end{equation}

where $\alpha,~\beta$ are dimensionless constants of order
unity.\\

We consider the Friedmann-Robertson-Walker (FRW) metric of the
universe as
\begin{equation}\label{2}
ds^{2}=-dt^{2}+a^{2}(t)\left( \frac{dr^{2}}{1-kr^{2}}+r^{2}\left(d
\theta^{2}+\sin^{2}\theta d \phi^{2}\right)\right)
\end{equation}

where $k=(0,\pm 1)$ is the curvature scalar and $a(t)$ is the
scale factor. Applying Clausius relation to apparent horizon of
the FRW universe with any spatial curvature and using the above
logarithmic corrected entropy area relation, Cai et al
\cite{Cai000} found the modified field equations \cite{Cai000,Akb}
in Einstein's gravity as

\begin{equation}\label{3}
\left(H^2+\frac{k}{a^2}\right)+\frac{\alpha
G}{2\pi}\left(H^2+\frac{k}{a^2}\right)^{2}-\frac{\beta
G^{2}}{3\pi^{2}}\left(H^2+\frac{k}{a^2}\right)^{3}=\frac{8\pi
G}{3}\rho
\end{equation}

and

\begin{equation}\label{4}
\left[1+\frac{\alpha
G}{\pi}\left(H^2+\frac{k}{a^2}\right)-\frac{\beta
G^{2}}{2\pi^{2}}\left(H^2+\frac{k}{a^2}\right)^{2}\right]
\left(\dot{H}-\frac{k}{a^2}\right)=-4\pi G(\rho+p)
\end{equation}

where $H=\frac{\dot{a}}{a}$ is the Hubble parameter. The bouncing
scenario and stability analysis for modified equations in
logarithmic corrected entropy has been investigated in \cite{Sal}.

\subsection{Modified Equations from Power Law Corrected Entropy}

From loop quantum gravity, the power law corrections arise due to
the entanglement of quantum fields moving into and out of the
horizon \cite{Das,Radicella,Sheykhi}. In this sense, the
entropy-area relation for power law correction can be given as

\begin{equation}
S=\frac{A}{4G}\left[1-K_{\epsilon}A^{1-\frac{\epsilon}{2}}\right],
\end{equation}

where,

\begin{equation}
K_{\epsilon}=\frac{\epsilon(4\pi)^{\frac{\epsilon}{2}-1}}{(4-\epsilon)r_{c}^{2-\epsilon}}
\end{equation}

Here, $r_{c}$ is the crossover scale and $\epsilon$ is a
dimensionless constant. By the similar procedure of Cai et al
\cite{Cai000} and applying Clausius relation to apparent horizon
of the FRW universe with a spatial curvature and using the above
power law corrected entropy area relation, one can obtain the
modified field equations for power law corrected entropy in
Einstein's gravity in the following form

\begin{equation}\label{7}
\left(H^2+\frac{k}{a^2}\right)-\frac{1}{{r_c}^{2-\epsilon}}
\left(H^2+\frac{k}{a^2}\right)^{\frac{\epsilon}{2}}=\frac{8\pi
G}{3}\rho
\end{equation}

and

\begin{equation}\label{8}
\left[2-\frac{\epsilon}{{r_c}^{2-\epsilon}}
\left(H^2+\frac{k}{a^2}\right)^{\frac{\epsilon}{2}-1}\right]
\left(\dot{H}-\frac{k}{a^2}\right)=-8\pi G(\rho+p)
\end{equation}

\section{A Brief Review of Ho$\check{\text r}$ava-Lifshitz Gravity Theory}

Inspired by Lifshitz theory in solid state physics, Horava
\cite{Horava1,Horava2,Horava3} proposed a UV complete field theory
of gravity which is a non-relativistic renormalizable theory of
gravity and reduces to Einstein's General Relativity at large
scales, referred to the Horava-Lifshitz (HL) gravity theory. In
the (3+1) dimensional Arnowitt-Deser-Misner (ADM) formalism, the
full metric can be written as \cite{Arn,Noji1}

\begin{equation}
ds^{2}=-N^{2}dt^{2} + g_{ij}(dx^{i} + N^{i}dt)(dx^{j} + N^{j}dt)
\end{equation}

Here the dynamical variables $N$ and $N_{i}$ are respectively the
lapse and shift functions. Under the detailed balance condition,
the full action for Ho$\check{\text r}$ava-Lifshitz gravity is
given by \cite{Hao}

\begin{eqnarray*}
S=\int dt d^{3}x \sqrt{g}N \left[\frac{2}{\kappa^{2}}(K_{ij}K^{ij}
- \lambda K^{2}) + \frac{\kappa^{2}}{2 \omega^{4}}C_{ij}C^{ij}-
\frac{\kappa^{2}\mu \epsilon^{ijk}}{2 \omega^{2} \sqrt{g}}
R_{il}\nabla_{j}R^{l}_{k}\right.
\end{eqnarray*}
\begin{equation}
\left.- \frac{\kappa^{2}\mu^{2}}{8}R_{ij}R^{ij}+
\frac{\kappa^{2}\mu^{2}}{8(1-3\lambda)}\left(\frac{1 -4
\lambda}{4}R^{2} + \Lambda R - 3 \Lambda^{2} \right)\right]
\end{equation}

where $K_{ij}$ is the extrinsic curvature taking the form:

\begin{equation}
K_{ij}=\frac{1}{2N}(\dot{g}_{ij} - \nabla_{i}N_{j}-
\nabla_{j}N_{i})
\end{equation}

and the Cotton tensor $C^{ij}$ takes the form:

\begin{equation}
C^{ij}=\frac{\epsilon^{ikl}}{\sqrt{g}}\nabla_{k}(R^{j}_{i} -
\frac{1}{4}R \delta^{j}_{l})
\end{equation}

Here dot denotes a derivative with respect to $t$ and the
covariant derivatives are defined with respect to the spatial
metric $g_{ij}$. Here $\epsilon^{ijk}$ is the totally
antisymmetric unit tensor, $\lambda$ is a dimensionless coupling
constant and the parameters $\kappa$, $\omega$ and $\mu$ are
constants with mass dimensions $-1,~ 0,~ 1$ respectively. Also
$\Lambda$ is related to the cosmological constant ($>0$) in the IR
limit. Now, in the cosmology, the projectability condition is
$N=1,~N^{i}=0$. By varying $N$ and $g_{ij}$, the non-vanishing
equations of motions can be expressed as

\begin{equation}
H^{2}=\frac{\kappa^{2}}{6(3\lambda -1)} ~ \rho +
\frac{\kappa^{2}}{6(3\lambda -1)} \left[\frac{3 \kappa^{2} \mu^{2}
k^{2}}{8(3\lambda -1)a^{4}} + \frac{3 \kappa^{2} \mu^{2}
\Lambda^{2}}{8(3\lambda -1)}\right] - \frac{ \kappa^{4} \mu^{2}k
\Lambda }{8(3\lambda -1)^{2}a^{2}}
\end{equation}

and

\begin{equation}
\dot{H} + \frac{3}{2}H^{2}= - \frac{\kappa^{2}}{4(3\lambda -1)}~p
- \frac{\kappa^{2}}{4(3\lambda -1)} \left[\frac{ \kappa^{2}
\mu^{2} k^{2}}{8(3\lambda -1)a^{4}}- \frac{3 \kappa^{2} \mu^{2}
\Lambda^{2}}{8(3\lambda -1)}\right] - \frac{ \kappa^{4} \mu^{2}k
\Lambda }{16(3\lambda -1)^{2}a^{2}}
\end{equation}

let us assume

\begin{equation}
G_{c}=\frac{\kappa^{2}}{16 \pi(3\lambda -1)}~,~~\nu=\frac{
\kappa^{4} \mu^{2} \Lambda }{8(3\lambda -1)^{2}}>0
\end{equation}

where $G_{c}$ is the ``cosmological" Newton's constant with
$\lambda>1/3$. In Einstein's gravity, the Newton's gravitational
constant is given by $G=\frac{\kappa^{2}}{32 \pi}$. So we can
write $G_{c}=\frac{2G}{3\lambda-1}$. In fact, for $\lambda=1$,
Lorentz invariance is restored ($G_{c}=G$), while for $\lambda \ne
1$, Lorentz invariance is broken ($G_{c}\ne G$). So we can express
the above Friedmann equations in the following form:

\begin{equation}\label{16}
H^{2} + \frac{k\nu}{a^{2}}=\frac{8\pi G_{c}}{3}~\rho +
\frac{k^{2}\nu}{2 \Lambda a^{4}} + \frac{\nu\Lambda}{2}
\end{equation}

and

\begin{equation}\label{17}
\dot{H} -\frac{k\nu}{a^{2}}= -4 \pi G_{c} (\rho+p) -
\frac{k^{2}\nu}{\Lambda a^{4}}
\end{equation}

\section{Dynamical System and Stability Analysis
for Entropy Corrected Models in Ho$\check{\text r}$ava-Lifshitz
Gravity}

\subsection{Logarithmic Corrected Model}

By the similar procedure of Cai et al \cite{Cai000} and applying
Clausius relation to apparent horizon of the FRW universe with a
spatial curvature and using the logarithmic corrected entropy area
relation, we can obtain the modified field equations in
Ho$\check{\text r}$ava-Lifshitz (HL) gravity. The modified field
equations for logarithmic corrected entropy will be the
modifications of the equations (\ref{3}), (\ref{4}), (\ref{16})
and (\ref{17}), which are presented in the following form:

\begin{equation}\label{fld eqn 1 HL log corr}
\left(H^2+\frac{k\nu}{a^2}\right)+\frac{\alpha
G_{c}}{2\pi}\left(H^2+\frac{k}{a^2}\right)^{2}-\frac{\beta
G_{c}^{2}}{3\pi^{2}}\left(H^2+\frac{k}{a^2}\right)^{3}=\frac{8\pi
G_{c}}{3}\rho+\frac{k^{2}\nu}{2 \Lambda a^{4}} +
\frac{\nu\Lambda}{2}
\end{equation}

and

\begin{equation}\label{fld eqn 2 HL log corr}
\left(\dot{H}-\frac{k\nu}{a^2}\right)+\left[\frac{\alpha
G_{c}}{\pi}\left(H^2+\frac{k}{a^2}\right)-\frac{\beta
G_{c}^{2}}{\pi^{2}}\left(H^2+\frac{k}{a^2}\right)^{2}\right]
\left(\dot{H}-\frac{k}{a^2}\right)=-4\pi G_{c}(\rho+p) -
\frac{k^{2}\nu}{\Lambda a^{4}}
\end{equation}

The continuity equation is given by

\begin{equation}
\dot{\rho}+3H(\rho+p)=0
\end{equation}

The study of dynamical system combined with phase space analysis
is considered to be a strong tool to understand the behavior of
the bouncing evolution of universe. It suggests that the
analytical solutions for the model needs to be stable (or
asymptotically stable under small perturbations). In the line with
this idea, we introduce the following variables

\begin{equation}
\chi=H,~~~~\zeta=a,~~~~\eta=\rho
\end{equation}

Then the field equations \eqref{fld eqn 1 HL log corr} and
\eqref{fld eqn 2 HL log corr} reduce to the following autonomous
system

\begin{equation}
\dot{\chi}=\frac{-4\pi G_c(1+\gamma)\eta-\frac{k^2\nu}{\Lambda
\zeta^4}+\frac{k}{\zeta^2}\left[(1+\nu)+\alpha'\left(\chi^2+\frac{k}{\zeta^2}\right)-\beta'{\left(\chi^2
+\frac{k}{\zeta^2}\right)}^2\right]}{1+\alpha'\left(\chi^2+\frac{k}{\zeta^2}\right)
-\beta'{\left(\chi^2+\frac{k}{\zeta^2}\right)}^2}
\end{equation}

and

\begin{equation}
\dot{\zeta}=\zeta\chi
\end{equation}

Here $\alpha'=\frac{\alpha G_c}{\pi}$ and $\beta'=\frac{\beta
{G_c}^2}{\pi^2}$. To find the critical points, we must have
$\dot{\chi}=0,\dot{\zeta}=0$. Solving them, the critical point
$(\chi_c,\zeta_c)$ can be obtained as

\begin{equation}
\chi_c=0
\end{equation}

and $\zeta_c$ is the root of the equation

\begin{equation}
\frac{3}{2}(1+\gamma)\nu\Lambda{\zeta_c}^6-3(1+\gamma)k\nu{\zeta_c}^4+\left[\frac{3}{2\Lambda}(1+\gamma)k^2\nu+k(1+\nu)
-(1+3\gamma)\alpha'k^2\right]{\zeta_c}^2=\frac{(1+3\gamma)}{2}\beta'k^3
\end{equation}

It can be observed easily that the critical points depend upon a
number of parameters. Thus one can have multiple analytical
solutions for the critical points by fine tuning the parameters.
Here we have obtained a stable critical point by assuming $k=-1$,
$\Lambda=\alpha'=\beta'=1$, $\gamma=1/3$, $\nu=1$. The graph of
$a$ vs $H$ in this case is shown in Fig.1 and we found the stable center.\\\\

\begin{figure}
~~~~~~~~~~~~~~~~~~~~~~~~~~~~~~~\includegraphics[height=3.0in]{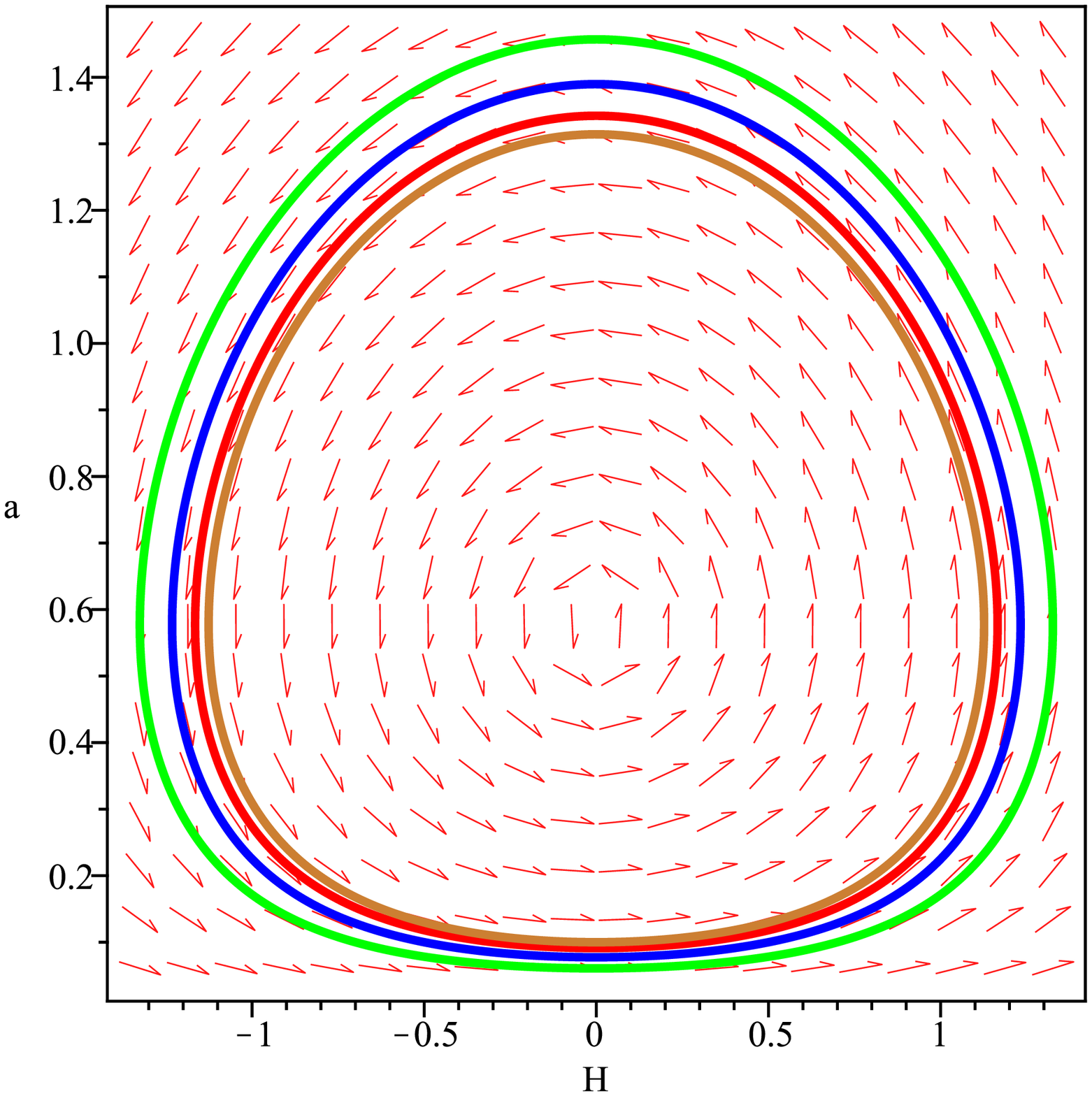}~~~~

\vspace{2mm} ~~~~~~~~~~~~~~~~~~~~~~~~~~~~~~~~~~~~~~~~~~~~~~~~~~~~~~~~~~~~~~~Fig.1~~~~~~~~~~~~\\
\vspace{4mm}

~~~~Fig.1: Phase diagram of $a$ vs $H$ in the Ho$\check{\text
r}$ava-Lifshitz gravity with logarithmic corrected model with the
choice of the parameters as
$k=-1$, $\Lambda=\alpha'=\beta'=1$, $\gamma=1/3$, $\nu=1$.\\

\vspace{6mm}

\end{figure}

\subsection{Power Law Corrected Model}

Similar to logarithmic corrected model in Ho$\check{\text
r}$ava-Lifshitz gravity, the modified field equations for power
law corrected entropy will be the modifications of the equations
(\ref{7}), (\ref{8}), (\ref{16}) and (\ref{17}), which are
presented in the following form:

\begin{equation}\label{fld eqn 1 HL power law corr}
\left(H^2+\frac{k\nu}{a^2}\right)-\frac{1}{{r_c}^{2-\epsilon}}
\left(H^2+\frac{k}{a^2}\right)^{\frac{\epsilon}{2}}=\frac{8\pi
G_{c}}{3}\rho+\frac{k^{2}\nu}{2 \Lambda a^{4}} +
\frac{\nu\Lambda}{2}
\end{equation}

and

\begin{equation}\label{fld eqn 2 HL power law corr}
2\left(\dot{H}-\frac{k\nu}{a^2}\right)-\frac{\epsilon}{{r_c}^{2-\epsilon}}
\left(H^2+\frac{k}{a^2}\right)^{\frac{\epsilon}{2}-1}
\left(\dot{H}-\frac{k}{a^2}\right)=-8\pi G_{c}(\rho+p) -
\frac{2k^{2}\nu}{\Lambda a^{4}}
\end{equation}

Similar to the case of logarithmic corrected model in HL gravity,
here we consider

\begin{equation}
\chi=H,~~~~\zeta=a,~~~~\eta=\rho
\end{equation}

Then the field equations \eqref{fld eqn 1 HL power law corr} and
\eqref{fld eqn 2 HL power law corr} reduce to the following
autonomous system

\begin{equation}
\dot{\chi}=\frac{\frac{2k\nu}{\zeta^2}-\frac{2k^2\nu}{\Lambda\zeta^4}-3(1+\gamma)\left[\left(\chi^2+\frac{k\nu}{\zeta^2}\right)
-\frac{1}{{r_c}^{2-\epsilon}}\left(\chi^2+\frac{k}{\zeta^2}\right)^{\frac{\epsilon}{2}}-\frac{k^2\nu}{2\Lambda\zeta^4}
-\frac{\nu\Lambda}{2}\right]-\frac{k\epsilon}{{r_c}^{2-\epsilon}\zeta^2}\left(\chi^2+\frac{k}{\zeta^2}\right)^{\frac{\epsilon}{2}-1}}
{2-\frac{\epsilon}{{r_c}^{2-\epsilon}}\left(\chi^2+\frac{k}{\zeta^2}\right)^{\frac{\epsilon}{2}-1}}
\end{equation}

and

\begin{equation}
\dot{\zeta}=\zeta\chi
\end{equation}

By putting $\dot{\chi}=0,\dot{\zeta}=0$, the critical point
$(\chi_c,\zeta_c)$ can be obtained as

\begin{equation}
\chi_c=0
\end{equation}

and $\zeta_c$ is the root of the equation

\begin{equation}
\frac{\nu\Lambda}{2}{\zeta_c}^{4-\epsilon}+\frac{k^{\frac{\epsilon}{2}}}{{r_c}^{2-\epsilon}}{\zeta_c}^{3-\epsilon}
+\frac{k^2\nu}{2\Lambda}{\zeta_c}^{-\epsilon}=\frac{\epsilon
k^{\frac{\epsilon}{2}}}{3(1+\gamma){r_c}^{2-\epsilon}}
\end{equation}

In this case as well, the critical solutions depend on a number of
parameters and hence more than one solution is possible by
adjusting the parameters involved. We have found a stable critical
point by choosing the parameters as $k=1$, $\Lambda=\nu=1$,
$\epsilon=2$, $r_c=10$ and $\gamma=1/3$. The phase space diagram
of $a$ vs $H$ is shown in Fig.2 and we found the stable center.\\\\

\begin{figure}
~~~~~~~~~~~~~~~~~~~~~~~~~~~~~~~\includegraphics[height=3.0in]{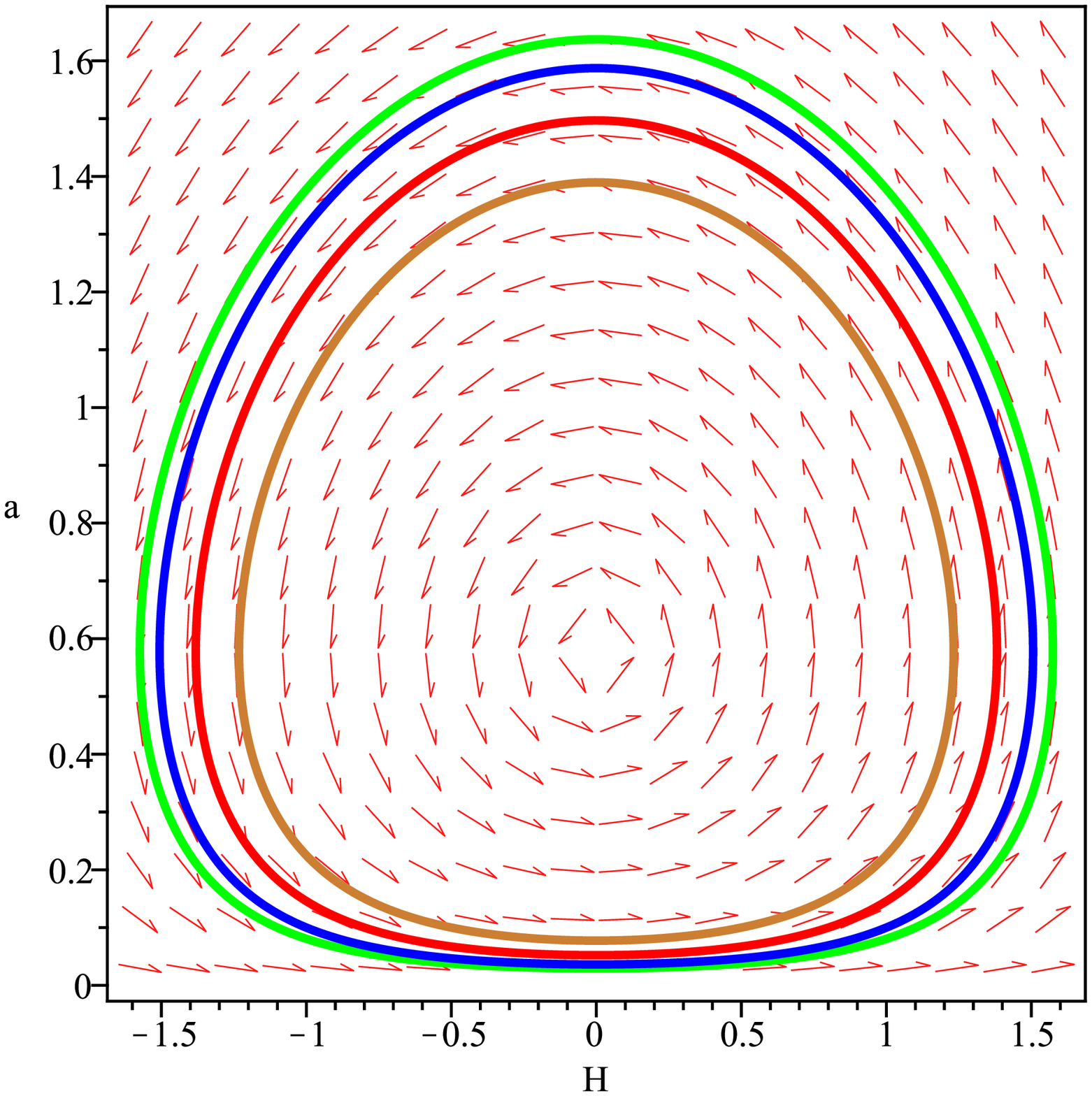}~~~~

\vspace{2mm} ~~~~~~~~~~~~~~~~~~~~~~~~~~~~~~~~~~~~~~~~~~~~~~~~~~~~~~~~~~~~~~~Fig.2~~~~~~~~~~~~\\
\vspace{4mm}

~~~~~Fig.2: Phase diagram of $a$ vs $H$ in Ho$\check{\text
r}$ava-Lifshitz gravity with power law corrected model with the
choice of the parameters involved as $k=1$,
$\Lambda=\nu=1$, $\epsilon=2$, $r_c=10$ and $\gamma=1/3$.\\

\vspace{6mm}

\end{figure}

\section{A Brief Review of Fractal Universe}

The idea of fractal effects in the Einstein's equations is one of
the approach to cosmic acceleration in the gravity theory. Fractal
features of quantum gravity and different cosmological properties
in fractal universe have been discussed by Calcagni
\cite{1GC10,Cal2}. In the Einstein's gravity, the action for the
effective fractal space-time is given by \cite{1GC10}

\begin{equation}\label{33}
S=\frac{1}{2\kappa^{2}}\int
 (d^{4}(x))v\sqrt{-g}~\left(R-2\Lambda-\omega\partial_{\mu}v\partial^{\mu}v+2\kappa^{2}{\cal L}_{m}\right)~,
\end{equation}

where $g=|g_{\mu\nu}|$, $\kappa^{2}=8\pi G$, $\Lambda$ is the
cosmological constant, $R$ is the Ricci scalar, $\omega$ is the
fractal parameter, $v$ is the fractal function and ${\cal L}_{m}$
is the matter Lagrangian. Taking the variation of the action given
in (\ref{33}), one can obtain the Friedmann equations in the
fractal universe as \cite{1GC10}

\begin{equation}\label{34}
\left(H^2+\frac{k}{a^2}\right)+H\frac{\dot{v}}{v}-\frac{\omega}{3}\dot{v}^{2}
=\frac{8\pi G}{3}\rho+\frac{\Lambda}{3}
\end{equation}

and

\begin{equation}\label{35}
2\left(\dot{H}-\frac{k}{a^2}\right)-H\frac{\dot{v}}{v}+\omega
\dot{v}^{2}+\frac{\ddot{v}}{v}=-8\pi G(\rho+p)
\end{equation}

The continuity equation is given by

\begin{equation}\label{conti eqn fractal}
\dot{\rho}+\left(3H+\frac{\dot{v}}{v}\right)(\rho+p)=0
\end{equation}

\section{Dynamical System and Stability Analysis for Entropy Corrected Models in Fractal Universe}

\subsection{Logarithmic Corrected Model}

Similar to logarithmic corrected model in Ho$\check{\text
r}$ava-Lifshitz gravity, the modified field equations
(modifications of the equations (\ref{34}), (\ref{35})) for
logarithmic corrected entropy in fractal Universe are obtained as

\begin{equation}\label{fld eqn 1 Fractal log corr}
\left(H^2+\frac{k}{a^2}\right)+\frac{\alpha
G}{2\pi}\left(H^2+\frac{k}{a^2}\right)^{2}-\frac{\beta
G^{2}}{3\pi^{2}}\left(H^2+\frac{k}{a^2}\right)^{3}+H\frac{\dot{v}}{v}-\frac{\omega}{3}\dot{v}^{2}
=\frac{8\pi G}{3}\rho+\frac{\Lambda}{3}
\end{equation}

and

\begin{equation}\label{fld eqn 2 Fractal log corr}
2\left[1+\frac{\alpha
G}{\pi}\left(H^2+\frac{k}{a^2}\right)-\frac{\beta
G^{2}}{\pi^{2}}\left(H^2+\frac{k}{a^2}\right)^{2}\right]
\left(\dot{H}-\frac{k}{a^2}\right)-H\frac{\dot{v}}{v}+\omega
\dot{v}^{2}+\frac{\ddot{v}}{v}=-8\pi G(\rho+p)
\end{equation}

We follow the same procedure as followed in the previous two cases
in Ho$\check{\text r}$ava-Lifshitz gravity. The variables are
defined in the similar manner as

\begin{equation}
\chi=H,~~~~\zeta=a,~~~~\eta=\rho
\end{equation}
Here we may choose $v=v_{0}a^n$, where $v_{0}$ and $n$ are
constants. Then the modified field equations \eqref{fld eqn 1
Fractal log corr} and \eqref{fld eqn 2 Fractal log corr} would
form the autonomous system given by

\begin{eqnarray*}
\dot{\chi}=\frac{1}{\left[1+\frac{n}{2}+\alpha'\left(\chi^2+\frac{k}{\zeta^2}\right)
-\beta'\left(\chi^2+\frac{k}{\zeta^2}\right)^2\right]}\left\{(1
+\gamma)\Lambda+\frac{2k}{\zeta^2}\left[1+\alpha'\left(\chi^2+\frac{k}{\zeta^2}\right)-\beta'\left(\chi^2
+\frac{k}{\zeta^2}\right)^{2}\right]\right.
\end{eqnarray*}
\begin{eqnarray*}
-\chi^2(-n+n^2+\omega
n^2{v_0}^2\zeta^{2n})-3(1+\gamma)\left[\left(\chi^2+\frac{k}{\zeta^2}\right)\left\{1+\frac{\alpha'}{2}\left(\chi^2
+\frac{k}{\zeta^2}\right)-\frac{\beta'}{3}\left(\chi^2+\frac{k}{\zeta^2}\right)^2\right\}\right.
\end{eqnarray*}
\begin{equation}
\left.\left.+\chi^2\left(n-\frac{1}{3}\omega
n^2{v_0}^2\zeta^{2n}\right)\right]\right\}
\end{equation}

and

\begin{equation}
\dot{\zeta}=\zeta\chi
\end{equation}

Here $\alpha'=\frac{\alpha G}{\pi}$ and $\beta'=\frac{\beta
G^2}{\pi^2}$. Solving the autonomous system, one can obtain the
critical point $(\chi_c,\zeta_c)$ as

\begin{equation}
\chi_c=0
\end{equation}

and $\zeta_c$ is the root of the equation

\begin{equation}
(1+\gamma)\Lambda{\zeta_c}^6-3(1+\gamma)k{\zeta_c}^4+\frac{(1-3\gamma)}{2}k^2\alpha'{\zeta_c}^2+\beta'\gamma
k^3=0
\end{equation}

Again, the analytical expression of the evolving equation for
$\zeta_c$ contains too many parameters. Hence we fine tune these
parameters and present a stable critical point for the following
choice of the parameters $k=1$, $\Lambda=\alpha'=\beta'=1$, $n=2$,
$v_0=0.2$, $\omega=0.3$ and $\gamma=0$. The phase diagram for the
stable center is drawn in Fig.3.\\\\

\begin{figure}
~~~~~~~~~~~~~~~~~~~~~~~~~~~~~~~\includegraphics[height=3.0in]{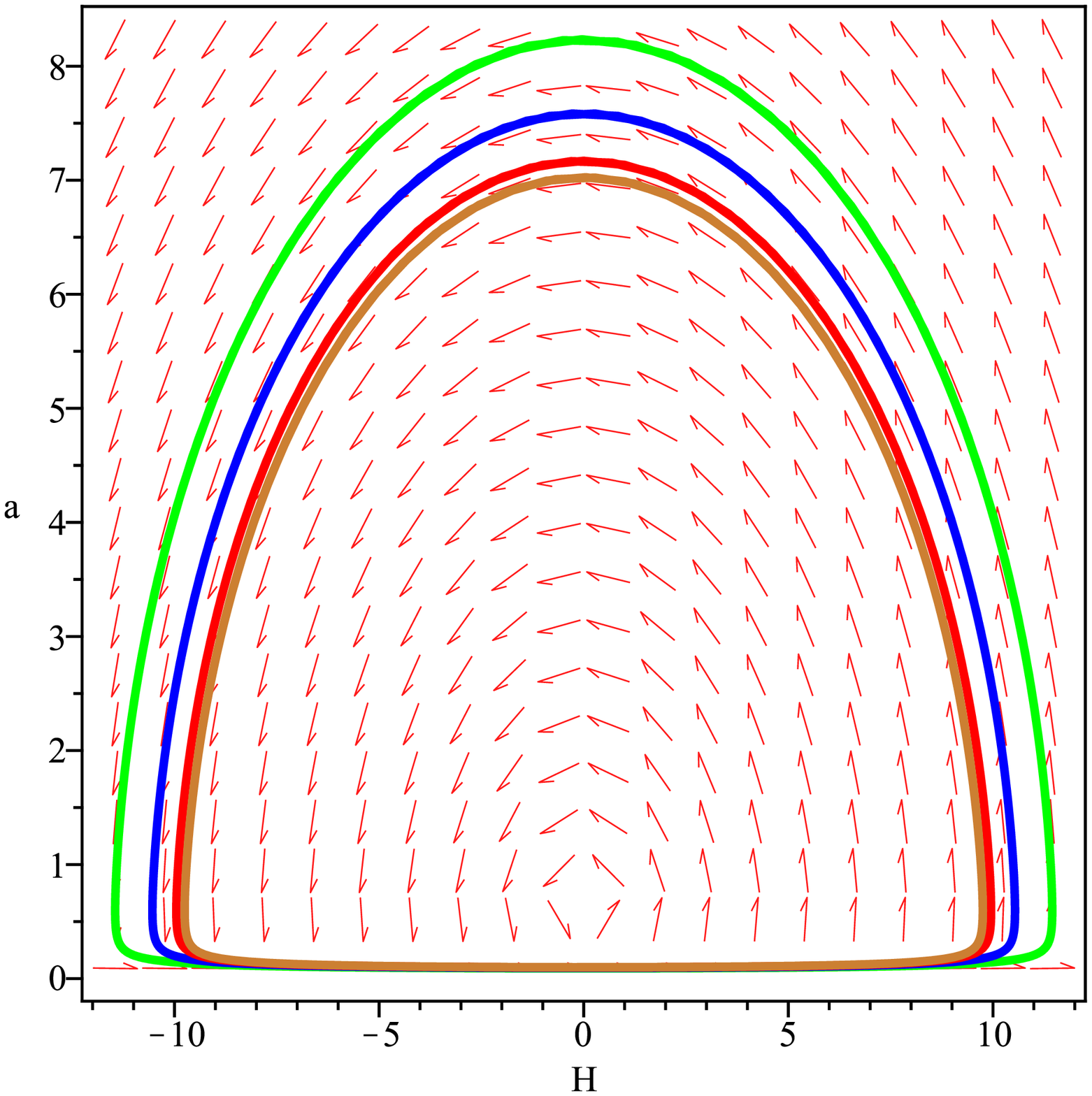}~~~~

\vspace{2mm} ~~~~~~~~~~~~~~~~~~~~~~~~~~~~~~~~~~~~~~~~~~~~~~~~~~~~~~~~~~~~~~~Fig.3~~~~~~~~~~~~\\
\vspace{4mm}

~~~~~Fig. 3: Phase diagram of $a$ vs $H$ in fractal universe with
logarithmic corrected model with the chosen values of the
parameters involved as $k=1$, $\Lambda=\alpha'=\beta'=1$,
$n=2,v_{0}=0.2,\omega=0.3$ and $\gamma=0$.\\

\vspace{6mm}

\end{figure}

\subsection{Power Law Corrected Model}

Similar to power law corrected model in Ho$\check{\text
r}$ava-Lifshitz gravity, the modified field equations
(modifications of the equations (\ref{34}), (\ref{35})) for power
law corrected entropy in fractal Universe are obtained as

\begin{equation}\label{fld eqn 1 Fractal power law corr}
\left(H^2+\frac{k}{a^2}\right)-\frac{1}{{r_c}^{2-\epsilon}}
\left(H^2+\frac{k}{a^2}\right)^{\frac{\epsilon}{2}}+H\frac{\dot{v}}{v}-\frac{\omega}{3}\dot{v}^{2}
=\frac{8\pi G}{3}\rho+\frac{\Lambda}{3}
\end{equation}

and

\begin{equation}\label{fld eqn 2 Fractal power law corr}
\left[2-\frac{\epsilon}{{r_c}^{2-\epsilon}}
\left(H^2+\frac{k}{a^2}\right)^{\frac{\epsilon}{2}-1}\right]
\left(\dot{H}-\frac{k}{a^2}\right)-H\frac{\dot{v}}{v}+\omega
\dot{v}^{2}+\frac{\ddot{v}}{v}=-8\pi G(\rho+p)
\end{equation}

A similar procedure is carried out with the variables defined as

\begin{equation}
\chi=H,~~~~\zeta=a,~~~~\eta=\rho
\end{equation}

Then the modified field equations \eqref{fld eqn 1 Fractal power
law corr} and \eqref{fld eqn 2 Fractal power law corr} reduce to
the following autonomous system

\begin{eqnarray*}
\dot{\chi}=\frac{1}{\left[2+n-\frac{\epsilon}{{r_c}^{2-\epsilon}}\left(\chi^2+\frac{k}{\zeta^2}\right)^{\frac{\epsilon}{2}-1}\right]}\left[\frac{2k}{\zeta^2}
-\frac{k\epsilon}{\zeta^2{r_c}^{2-\epsilon}}\left(\chi^2+\frac{k}{\zeta^2}\right)^{\frac{\epsilon}{2}-1}\right.
\end{eqnarray*}
\begin{equation}
\left.+\chi^2(n-n^2-n^2\omega{v_0}^2\zeta^{2n})-3(1+\gamma)\left\{\frac{k}{\zeta^2}-\frac{\Lambda}{3}
+\chi^2\left(1+n-\frac{\omega}{3}n^2{v_0}^2\zeta^{2n}\right)-\frac{1}{{r_c}^{2-\epsilon}}\left(\chi^2+\frac{k}{\zeta^2}\right)^{\frac{\epsilon}{2}}\right\}\right]
\end{equation}

and

\begin{equation}
\dot{\zeta}=\zeta\chi
\end{equation}

The critical point $(\chi_c,\zeta_c)$ can be obtained by solving
the above equations as

\begin{equation}
\chi_c=0
\end{equation}

and $\zeta_c$, beings the root of the equation

\begin{equation}
(3+3\gamma-\epsilon)k^{\frac{\epsilon}{2}}\left(\frac{\zeta_c}{r_c}\right)^{2-\epsilon}
=k(1+3\gamma)-\Lambda(1+\gamma){\zeta_c}^2
\end{equation}

Again we choose particular values to the various parameters in
order to attain a stable critical point. The values are $k=-1$,
$\Lambda=1$, $\epsilon=4$, $r_c=10$, $n=2, v_{0}=0.5, \omega=0.6$
and $\gamma=-1/2$. The phase space diagram
of $a$ vs $H$ is presented in Fig.4 and we found the stable center.\\\\

\begin{figure}
~~~~~~~~~~~~~~~~~~~~~~~~~~~~~~~\includegraphics[height=3.0in]{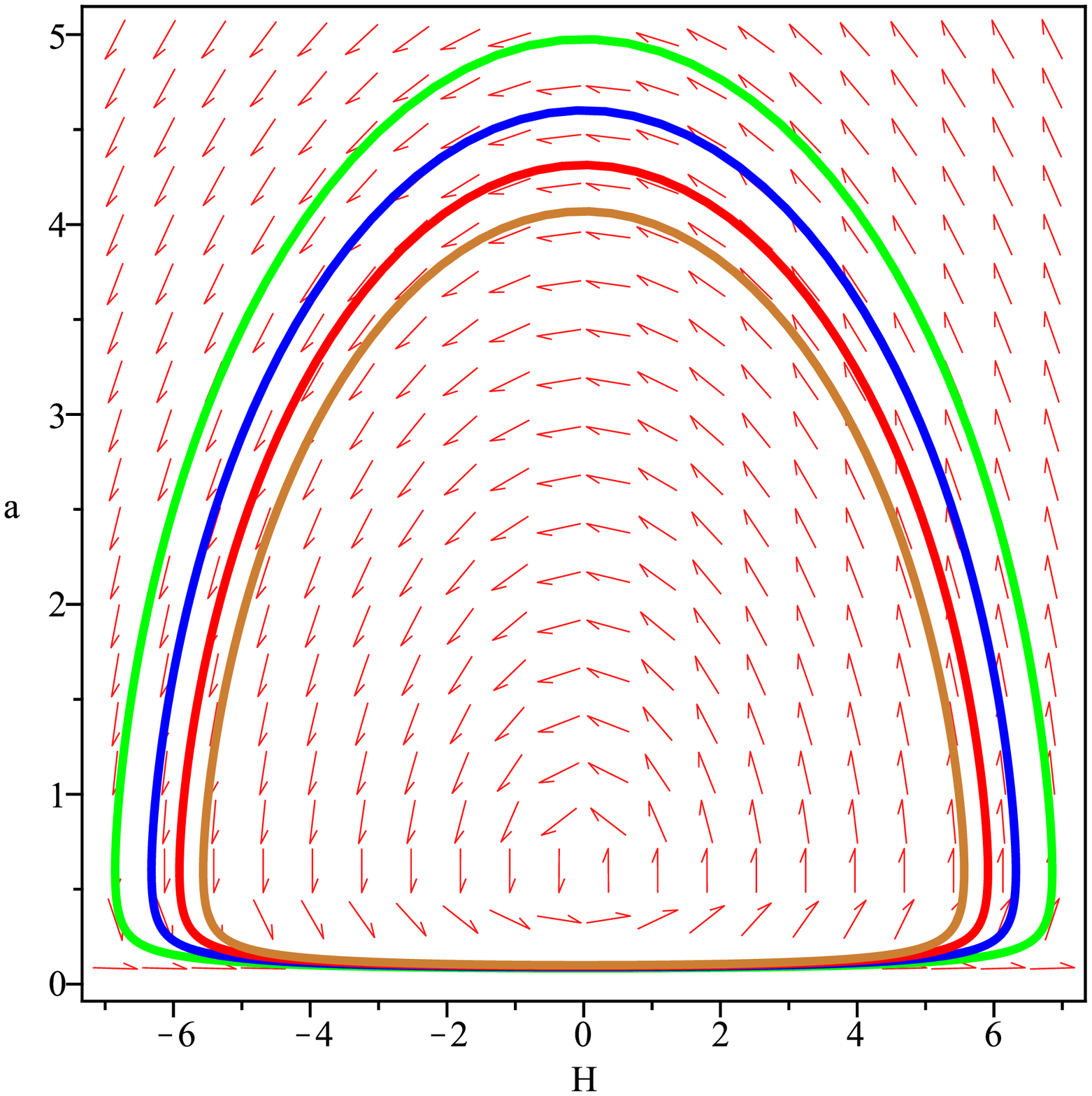}~~~~

\vspace{2mm} ~~~~~~~~~~~~~~~~~~~~~~~~~~~~~~~~~~~~~~~~~~~~~~~~~~~~~~~~~~~~~~~Fig.4~~~~~~~~~~~~\\
\vspace{4mm}

~~~~~Fig. 4: Phase diagram of $a$ vs $H$ in fractal universe with
power law corrected model with the chosen values of the parameters
involved as $k=-1$, $\Lambda=1$, $\epsilon=4$, $r_c=10$, $n=2,
v_{0}=0.5, \omega=0.6$ and $\gamma=-1/2$.\\

\vspace{6mm}

\end{figure}

\section{Connections Between Bounce and Energy Conditions}

To avoid the initial singularity problem of General Relativity,
many attempts had been made by proposing different modified or
generalized theories of gravity. One of these is the non-singular
bouncing cosmological model. By modifying the action, one can
achieve an oscillating universe with an initial narrow state by a
minimal radius, which further develops to an expanding phase. This
implies that the universe arrives into the Big Bang era after the
bounce happens, thus avoids the initial singularity and results in
a cyclic evolution. In terms of the scale factor, one can make the
argument as follows. During the initial state of bounce, the scale
factor decreases ($\dot{a}(t)<0$) and the universe gets
contracted. At the time of `bounce' ($t = t_{b}$, say), the scale
factor attains a minimum. The criteria for this is $\dot{a}(t)=0$
and $\ddot{a}(t)>0$ for $t\in (t_{b}-\epsilon, t_{b})\cup
(t_{b},t_{b}+\epsilon)$ for small $\epsilon>0$. After the bounce,
the scale factor increases again ($\dot{a}(t)>0$) and subsequently
the universe undergoes an expanding phase. One must note that for
a non-singular bounce, $a(t_{b})\ne 0$. However, the conditions
may not be sufficient.\\

In this section, we investigate the bouncing solutions of the
present models in the light of energy conditions. Our main focus
is to verify the Null Energy Condition (NEC) ($\rho+p\geq0$) as
violation of this condition automatically implies the violation of
all other point-wise energy conditions.\\

\subsection{Logarithmic Corrected Model in HL Gravity}

Here we use \eqref{fld eqn 2 HL log corr} during bounce and have

\begin{equation}\label{NEC Check HL Log}
4\pi{G_c}(\rho_b+p_b)=\left(\frac{k\nu}{{a_b}^2}+\frac{\alpha'k^2}{{a_b}^4}-\frac{k^2\nu}{\Lambda{a_b}^4}\right)
-\frac{\ddot{a_b}}{a_b}\left(1+\frac{\alpha'k}{{a_b}^2}-\frac{\beta'k^2}{{a_b}^4}\right)
\end{equation}

Here, $\rho_b,~p_b$ and $a_b$ signify the energy density, pressure
and the scale factor during bounce, i.e., at $t=t_{b}$. The right
hand side of the above equation must be positive for NEC to be
satisfied. The energy density at bounce in this case is given by

\begin{equation}
\rho_b=\frac{3}{8\pi{G_c}}\left[\frac{k\nu}{{a_b}^2}+\frac{\alpha'k^2}{{a_b}^4}-\frac{k^2\nu}{2\Lambda{a_b}^4}
-\frac{\beta'k^3}{3{a_b}^6}-\frac{\nu\Lambda}{2}\right]
\end{equation}

At the bounce, the continuity equation

\begin{equation}
\dot{\rho}=-3H(\rho+p)
\end{equation}

would become

\begin{equation}
\dot{\rho_b}=-3H_b(\rho_b+p_b)
\end{equation}

As we know that $H_b=\frac{\dot{a_b}}{a_b}=0$, hence
$\dot{\rho_b}=0$ and consequently $\rho_b$ reaches an extremum
energy density at bounce. Further

\begin{equation}\label{continuity in HL Log corr}
\ddot{\rho_b}=-3\dot{H_b}(\rho_b+p_b)
\end{equation}

Setting different values of $k$, we can find the restrictions for
the satisfaction of NEC and consequently the extremum energy
density during bounce.

\subsubsection{Case I: $k=0$}

From \eqref{NEC Check HL Log}, we get

\begin{equation}
(\rho_b+p_b)=-\frac{1}{4\pi{G_c}}\left(\frac{\ddot{a_b}}{a_b}\right)<0,~~
\text{since}~~\ddot{a_b}>0
\end{equation}

It can be seen that the NEC is thus being violated. Also we see
that $\ddot{\rho_b}>0$. Therefore $\rho_b$ must be attaining a
minimum value during bounce given by
$(\rho_{b})_{min}=-\frac{3\nu\Lambda}{16\pi{G_c}}$. This case
therefore turns out to be unfavorable for bounce.

\subsubsection{Case II: $k=\pm1$}

From \eqref{continuity in HL Log corr} we can say that in order to
satisfy the NEC, we must have $\ddot{\rho_b}<0$ as $\dot{H_b}>0$
during bounce. Thus $\rho_b$ in both the cases of $k=1$ and $k=-1$
attains a maximum value given by

\begin{equation}\label{rho at k=1}
{\rho_b}_{\text{max}}|_{k=1}=\frac{3}{8\pi{G_c}}\left[\frac{\nu}{{a_b}^2}+\left(\alpha'-\frac{\nu}{2\Lambda}\right)\frac{1}{{a_b}^4}
-\frac{\beta'}{3{a_b}^6}-\frac{\nu\Lambda}{2}\right]
\end{equation}

\begin{equation}\label{rho at k=-1}
{\rho_b}_{\text{max}}|_{k=-1}=\frac{3}{8\pi{G_c}}\left[-\frac{\nu}{{a_b}^2}+\left(\alpha'-\frac{\nu}{2\Lambda}\right)\frac{1}{{a_b}^4}
+\frac{\beta'}{3{a_b}^6}-\frac{\nu\Lambda}{2}\right]
\end{equation}

The criteria for the satisfaction of NEC for the two cases are

\begin{equation}\label{cond at k=1}
\frac{\alpha'}{{a_b}^2}+\nu\left(1-\frac{1}{\Lambda{a_b}^2}\right)>{a_b}
\ddot{a_b}\left(1+\frac{\alpha'}{{a_b}^2}-\frac{\beta'}{{a_b}^4}\right),~~~k=1
\end{equation}

\begin{equation}\label{cond at k=-1}
\frac{\alpha'}{{a_b}^2}-\nu\left(1+\frac{1}{\Lambda{a_b}^2}\right)>{a_b}\ddot{a_b}
\left(1-\frac{\alpha'}{{a_b}^2}-\frac{\beta'}{{a_b}^4}\right),~~~k=-1
\end{equation}

The evolution of the scale factor $a$, the sum of the energy
density and pressure $(\rho+p)$ and the energy density $\rho$
during the bounce period for logarithmic corrected universe in HL
gravity are shown in Figs. 5-7. Variations in the initial
conditions result in different colored plots. A barotropic
equation of state $p=\gamma\rho$ is assumed for the matter
content. For viable choices of the parameters involved, the plots
show an oscillating universe with the NEC being satisfied.\\

\begin{figure}
\vspace{4mm}

\includegraphics[height=2.0in]{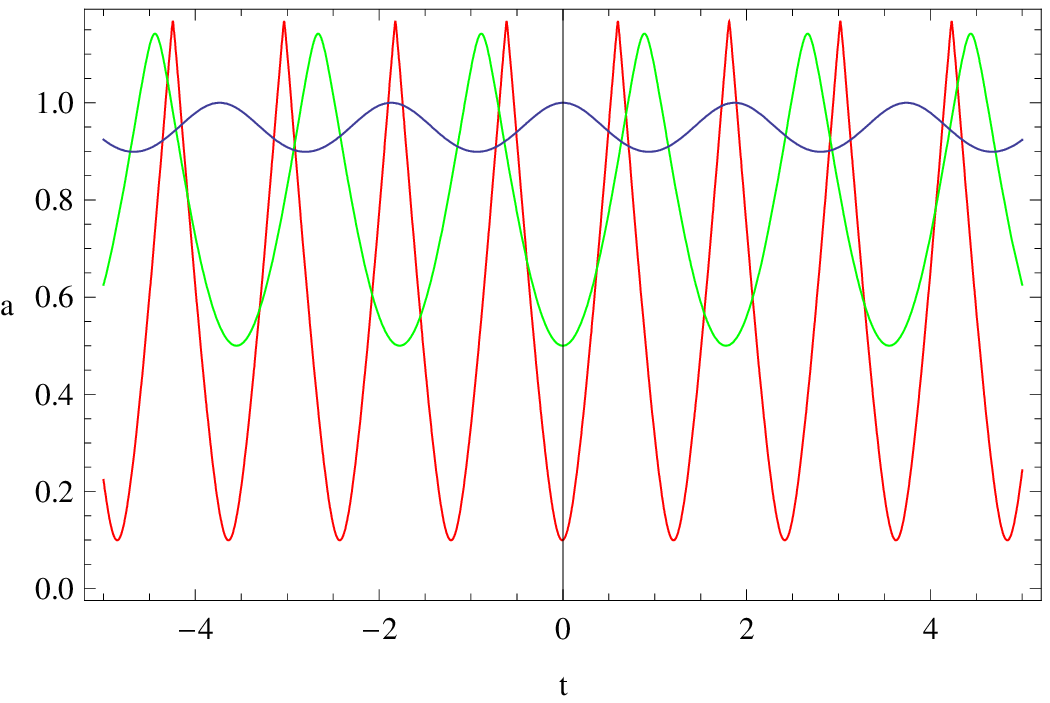}~~~~
\includegraphics[height=2.0in]{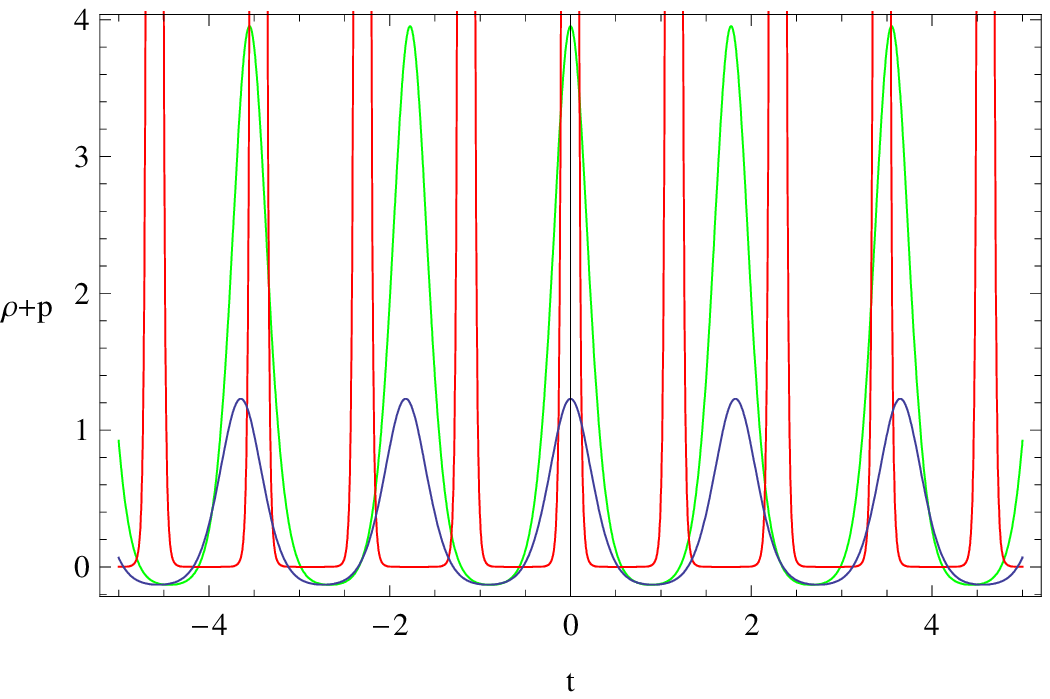}\\
\vspace{2mm} ~~~~~~~~~~~~~~~~~~~~~~~~~~~~~Fig.5~~~~~~~~~~~~~~~~~~~~~~~~~~~~~~~~~~~~~~~~~~~~~~~~~~~~~~~~~~~Fig.6~~~~\\
\vspace{4mm}

~~~~~~~~~~~~~~~~~~~~~~~~~~~~~~~~~~~~~\includegraphics[height=2.0in]{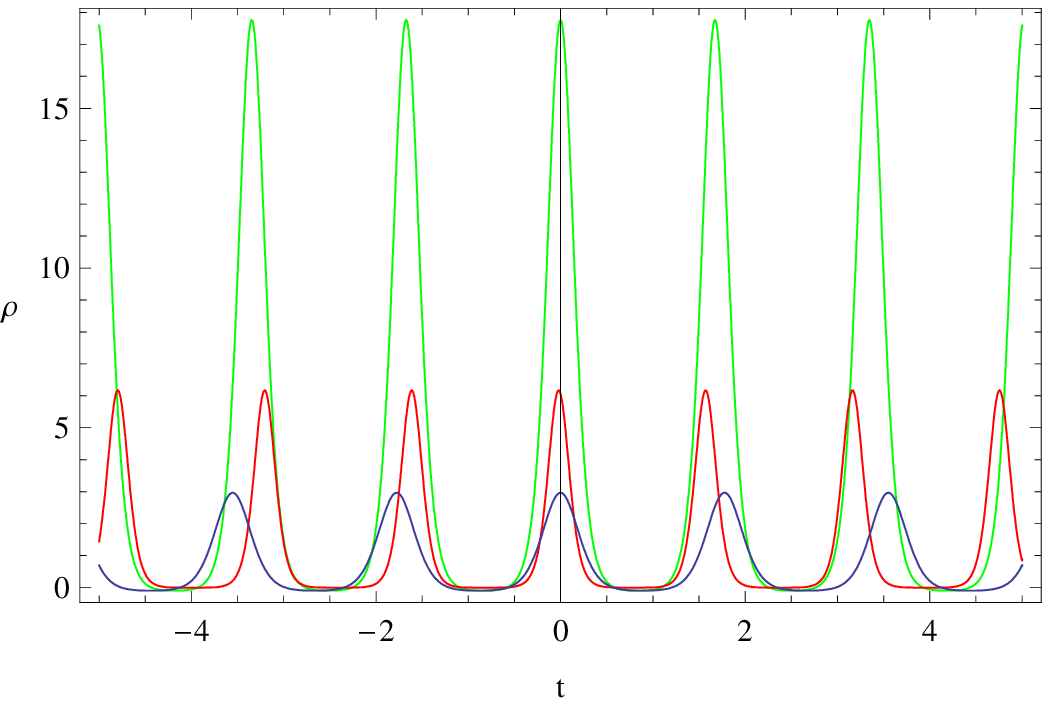}\\
\vspace{2mm} ~~~~~~~~~~~~~~~~~~~~~~~~~~~~~~~~~~~~~~~~~~~~~~~~~~~~~~~~~~~~~~~~~Fig.7~~~~~~~~~~~~~~\\

~~~Figs. 5-7 show the dynamical behaviors of the scale factor $a$,
the sum of the energy density and pressure $(\rho+p)$ and the
energy density $\rho$ during the bounce period for logarithmic
corrected universe in HL gravity. The parameters are chosen for
this case as $k=-1$, $G_c=\alpha'=\beta'=\nu=\Lambda=1$ and
$\gamma=1/3$.
Different colors signify different initial conditions.\\

\vspace{4mm}

\end{figure}

\subsection{Power Law Corrected Model in HL Gravity}

From \eqref{fld eqn 2 HL power law corr} we have

\begin{equation}\label{NEC Check HL Power law}
8\pi{G_c}(\rho_b+p_b)=\left[\frac{2k\nu}{{a_b}^2}\left(1-\frac{k}{\Lambda{a_b}^2}\right)-\frac{\epsilon
k^{\frac{\epsilon}{2}}}{{a_b}{r_c}^{2-\epsilon}}\right]-\frac{\ddot{a_b}}{a_b}\left[2-\frac{\epsilon}{{r_c}^{2-\epsilon}}
\left(\frac{k}{{a_b}^2}\right)^{\frac{\epsilon}{2}-1}\right]
\end{equation}

For the NEC to be satisfied, the right hand side of the above
equation needs to be positive. The energy density at the time of
bounce in this case is given by

\begin{equation}
\rho_b=\frac{3}{8\pi{G_c}}\left[\frac{k\nu}{{a_b}^2}-\frac{1}{{r_c}^{2-\epsilon}}
\left(\frac{k}{{a_b}^2}\right)^\frac{\epsilon}{2}
-\frac{k^2\nu}{2\Lambda{a_b}^4}-\frac{\nu\Lambda}{2}\right]
\end{equation}

\subsubsection{Case I: $k=0$}

This case is identical to the subsection 7.1.1. Here also the NEC
is violated and at the time of bounce, the energy density becomes
negative concluding no bounce.

\subsubsection{Case II: $k=\pm1$}

From \eqref{continuity in HL Log corr} we can say that in order to
satisfy the NEC, we get $\ddot{\rho_b}<0$. Thus $\rho_b$ in both
the cases of $k=1$ and $k=-1$ attains a maximum value given by

\begin{equation}
{\rho_b}_{\text{max}}|_{k=1}=\frac{3}{8\pi{G_c}}\left[\frac{\nu}{{a_b}^2}-\frac{\nu}{2\Lambda{a_b}^4}
-\frac{\nu\Lambda}{2}-\frac{1}{{a_b}^\epsilon{r_c}^{2-\epsilon}}\right]
\end{equation}

\begin{equation}\label{case k=-1}
{\rho_b}_{\text{max}}|_{k=-1}=-\frac{3}{8\pi{G_c}}\left[\frac{\nu}{{a_b}^2}+\frac{\nu}{2\Lambda{a_b}^4}
+\frac{\nu\Lambda}{2}+\frac{(-1)^{\frac{\epsilon}{2}}}{{a_b}^\epsilon{r_c}^{2-\epsilon}}\right]
\end{equation}

It is interesting to note that the fourth term of \eqref{case
k=-1} is the deciding term for a feasible $\rho_b$ during bounce.
For $\epsilon=2m$, where $m$ is odd, there would be a chance to
have a positive $\rho_b$. In all other cases, $\rho_b$ will be
negative concluding unfavorable condition for bounce. The criteria
for the satisfaction of NEC for $k=1$ and $k=-1$ will be

\begin{equation}
\frac{2\nu}{a_b}\left(1-\frac{1}{\Lambda{a_b}^2}\right)-\frac{\epsilon}{{r_c}^{2-\epsilon}}
>\ddot{a_b}\left[2-\frac{\epsilon}{({a_b}/{r_c})^{\epsilon-2}}\right],~~~k=1
\end{equation}

\begin{equation}
\frac{2\nu}{a_b}\left(1+\frac{1}{\Lambda{a_b}^2}\right)+\frac{\epsilon(-1)^{\frac{\epsilon}{2}}}{{r_c}^{2-\epsilon}}
+\ddot{a_b}\left[2+\frac{\epsilon(-1)^{\frac{\epsilon}{2}}}{({a_b}/{r_c})^{\epsilon-2}}\right]>0,~~~k=-1
\end{equation}

The evolution of the scale factor $a$, the sum of the energy
density and pressure $(\rho+p)$ and the energy density $\rho$
during the bounce period for power law corrected universe in HL
gravity are shown in Figs. 8-10. A barotropic equation of state
$p=\gamma\rho$ is assumed for the matter content. Different sets
of parameters have been chosen to generate the plots. For scale
factor, the parameters involved are chosen as
$k=\epsilon=\nu=\Lambda=G_c=1$ and $r_c=10$ to observe a bouncing
solution in this model. Here the variation is considered in the
initial conditions which produces different colored plots. On the
other hand, to generate the plots of $(\rho+p)$ and $\rho$, the
choice of $\epsilon$ is varied since this plays a major role in
defining the satisfaction of the NEC. Here we chose $k=-1$,
$\nu=\Lambda=G_c=1$, $r_c=10$ and $\epsilon=6$ (green line), $10$
(red line) and $14$ (blue line). it can be observed that the plots
show the satisfaction of the NEC.

\begin{figure}
\vspace{4mm}

\includegraphics[height=2.0in]{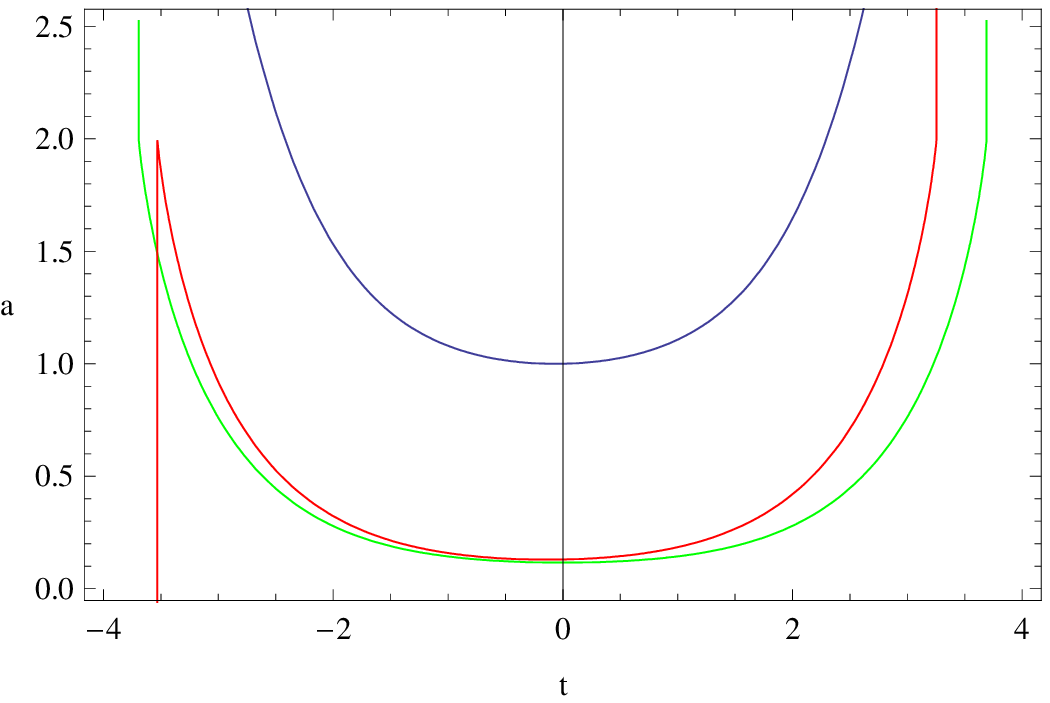}~~~~
\includegraphics[height=2.0in]{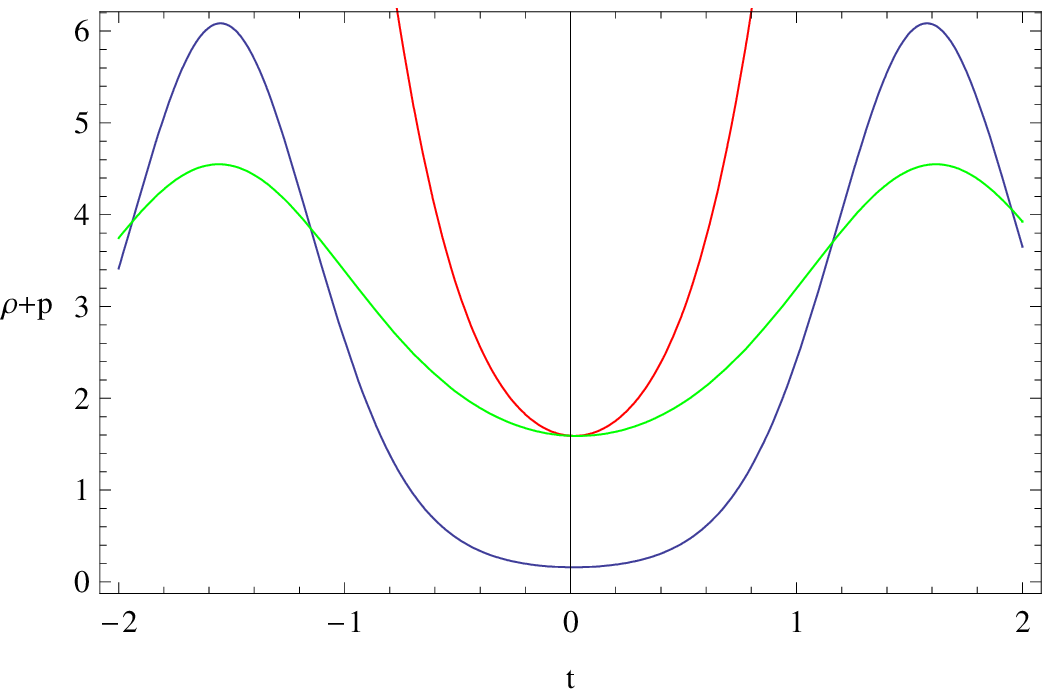}\\
\vspace{2mm} ~~~~~~~~~~~~~~~~~~~~~~~~~~~~~Fig.8~~~~~~~~~~~~~~~~~~~~~~~~~~~~~~~~~~~~~~~~~~~~~~~~~~~~~~~~~~~~~~~~Fig.9~~~~\\
\vspace{4mm}

~~~~~~~~~~~~~~~~~~~~~~~~~~~~~~~~~~~~~\includegraphics[height=2.0in]{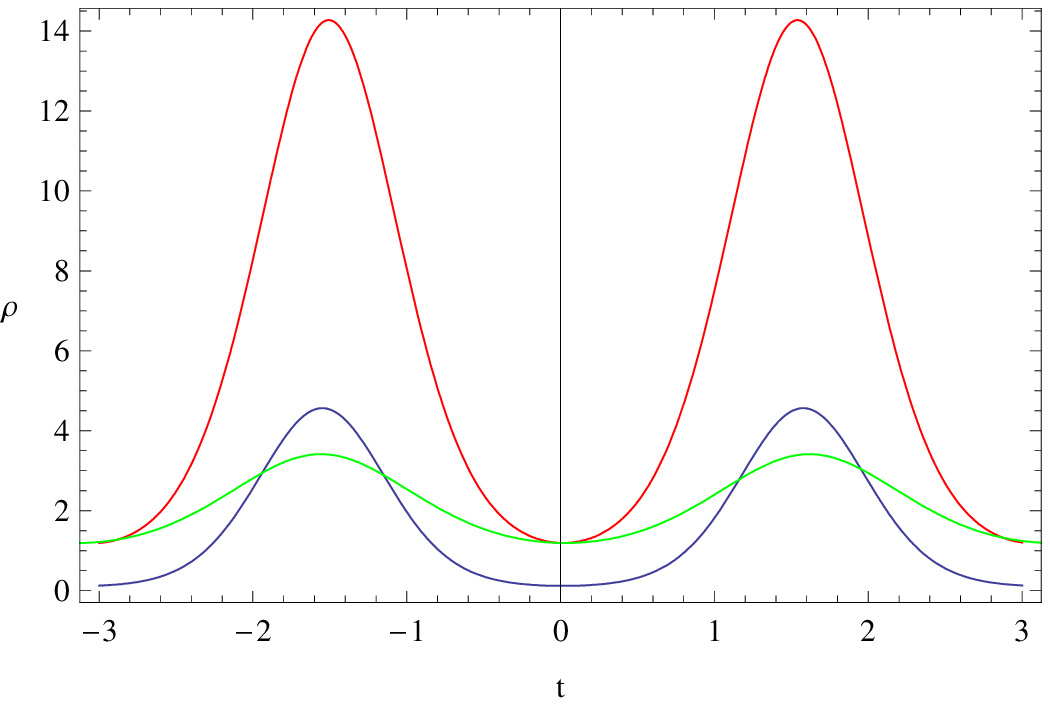}\\
\vspace{2mm} ~~~~~~~~~~~~~~~~~~~~~~~~~~~~~~~~~~~~~~~~~~~~~~~~~~~~~~~~~~~~~~~~~Fig.10~~~~~~~~~~~~~~\\

~~~~Figs. 8-10 show the dynamical behaviors of the scale factor
$a$, the sum of the energy density and pressure $(\rho+p)$ and the
energy density $\rho$ during the bounce period for power law
corrected universe in HL gravity.\\

\vspace{4mm}

\end{figure}

\subsection{Logarithmic Corrected Model in Fractal Universe}

From \eqref{fld eqn 2 Fractal log corr} we get

\begin{equation}
8\pi
G(\rho_b+p_b)=\frac{2k^2}{{a_b}^4}\left(1+\alpha'-\frac{\beta'
k}{{a_b}^2}\right)-\frac{\ddot{a_b}}{a_b}\left(n+\frac{2k\alpha'}{{a_b}^2}\right)
\end{equation}

where the energy density at the time of bounce in this case is
given by

\begin{equation}
\rho_b=\frac{3}{8\pi
G}\left[\frac{k}{{a_b}^2}+\frac{k^2\alpha'}{{a_b}^4}-\frac{k^3\beta'}{3{a_b}^6}
-\frac{\Lambda}{3}\right]
\end{equation}

From the continuity equation \eqref{conti eqn fractal}, we can
write

\begin{equation}
\dot{\rho_b}=-(n+3)H_b(\rho_b+p_b)
\end{equation}

Again as $H_b=0$, we have $\dot{\rho_b}=0$ and subsequently
$\rho_b$ attains a local extremum during bounce. Further

\begin{equation}\label{continuity in Fractal universe}
\ddot{\rho_b}=-(n+3)\dot{H_b}(\rho_b+p_b)
\end{equation}

Assuming the validity of NEC, we therefore would have
$\ddot{\rho_b}<0$ if $n>-3$ implying a local maximum value for
$\rho_b$ and otherwise a local minimum value for $\rho_b$.

\subsubsection{Case I: $k=0$}

In this case we have

\begin{equation}
\rho_b+p_b=-\frac{n}{8\pi G}\frac{\ddot{a_b}}{a_b}
\end{equation}

It is evident that the NEC is satisfied for a negative $n$ and
violated otherwise. The energy density at bounce is given by
$\rho_b=-\frac{\Lambda}{8\pi G}<0$. Thus this case is also not
favorable for bounce.

\subsubsection{Case II: $k=\pm1$}

In this case the criteria for the satisfaction of NEC are given by

\begin{equation}
2\left(1+\alpha'-\frac{\beta'}{{a_b}^2}\right)>a_b\ddot{a_b}\left(n+\frac{2\alpha'}{{a_b}^2}\right),~~~k=1
\end{equation}

and

\begin{equation}
2\left(1+\alpha'+\frac{\beta'}{{a_b}^2}\right)>a_b\ddot{a_b}\left(n-\frac{2\alpha'}{{a_b}^2}\right),~~~k=-1
\end{equation}

The evolution of the scale factor $a$, the sum of the energy
density and pressure $(\rho+p)$ and the energy density $\rho$
during the bounce period for logarithmic corrected model in
fractal universe are shown in Figs. 11-13. A barotropic equation
of state $p=\gamma\rho$ is assumed for the matter content. Here
the initial conditions are kept constant but the choice of
$\gamma$ is varied. For viable choices of the parameters involved,
the plots clearly show a bouncing solution with the NEC being
satisfied.

\begin{figure}
\vspace{4mm}

\includegraphics[height=2.0in]{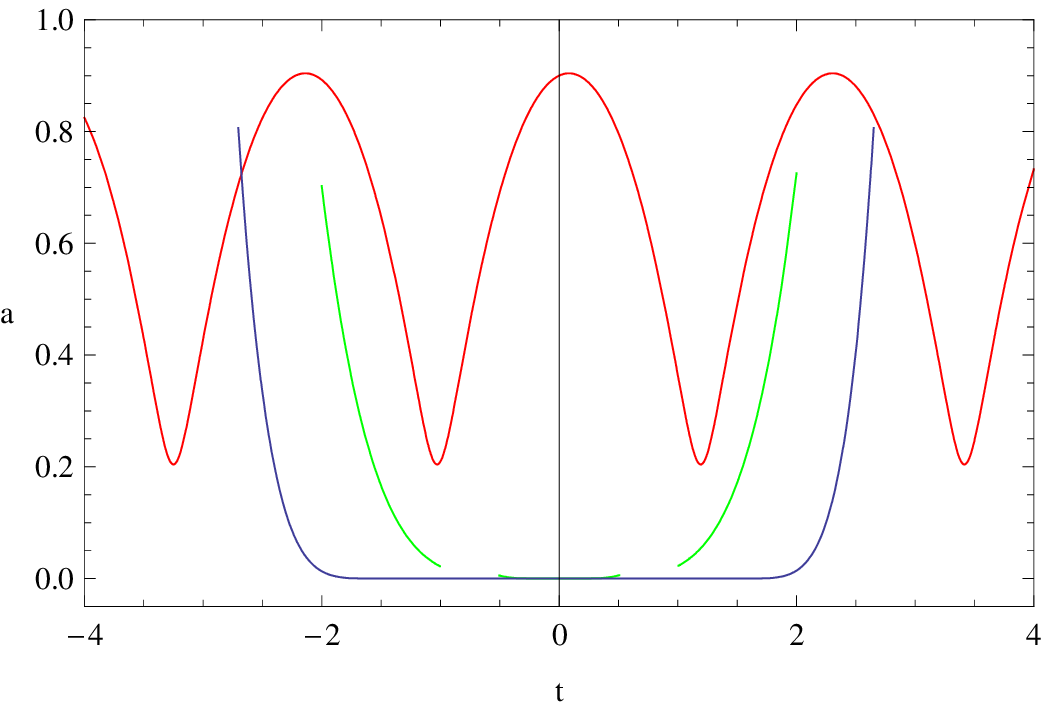}~~~~
\includegraphics[height=2.0in]{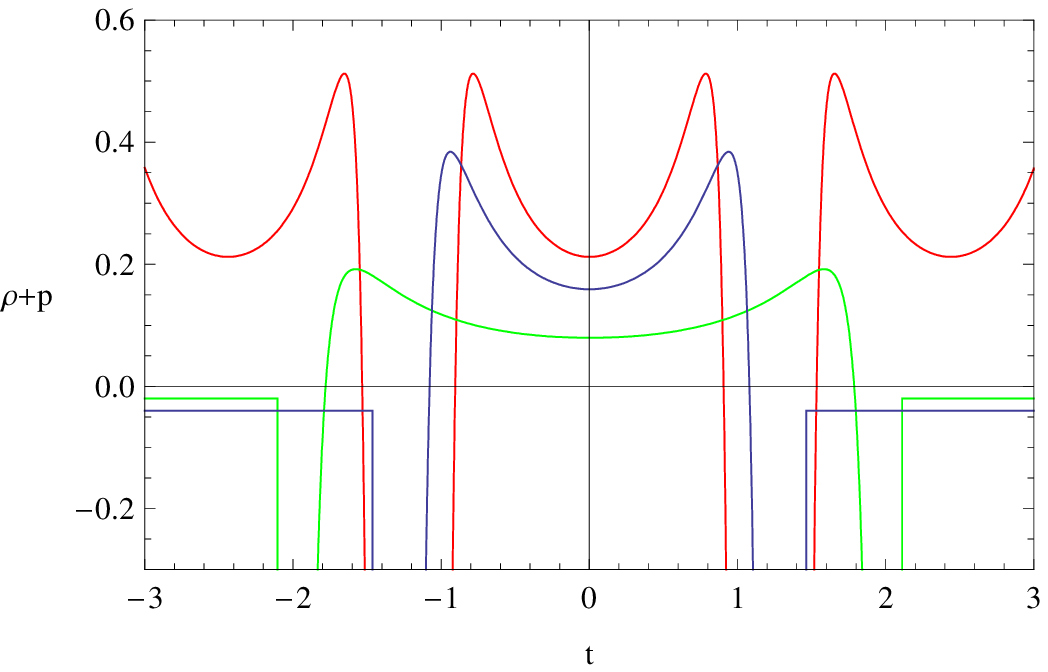}\\
\vspace{2mm} ~~~~~~~~~~~~~~~~~~~~~~~~~~~~~Fig.11~~~~~~~~~~~~~~~~~~~~~~~~~~~~~~~~~~~~~~~~~~~~~~~~~~~~~~~~~~~Fig.12~~~~\\
\vspace{4mm}

~~~~~~~~~~~~~~~~~~~~~~~~~~~~~~~~~~~~~\includegraphics[height=2.0in]{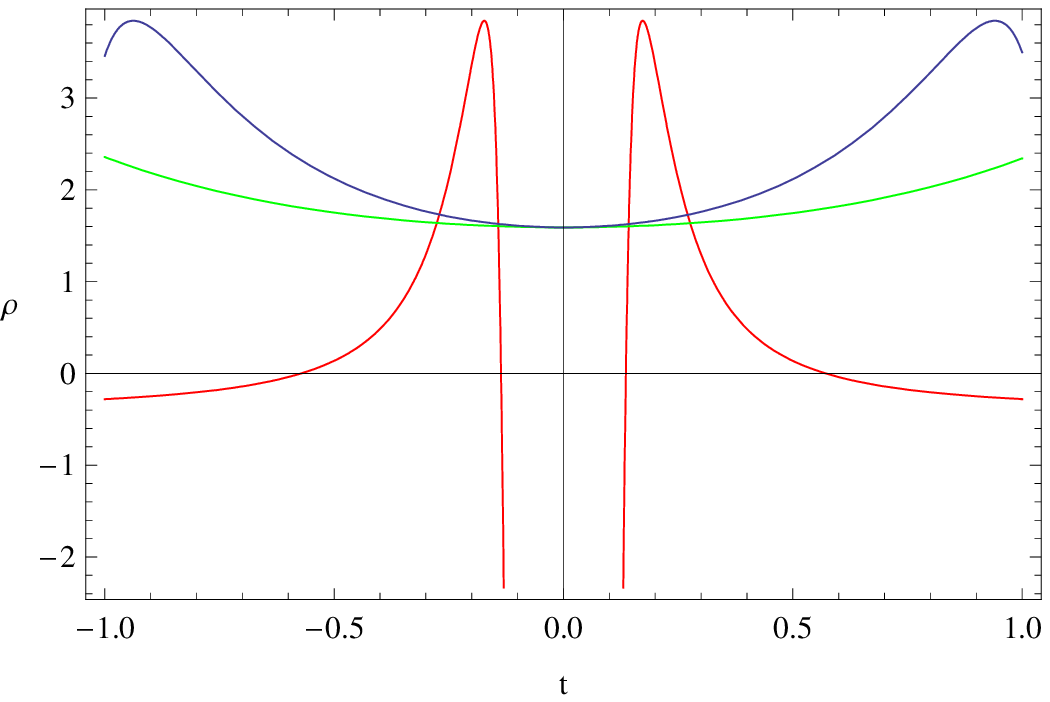}\\
\vspace{2mm} ~~~~~~~~~~~~~~~~~~~~~~~~~~~~~~~~~~~~~~~~~~~~~~~~~~~~~~~~~~~~~~~~~Fig.13~~~~~~~~~~~~~~\\

~~~Figs. 11-13 show the dynamical behaviors of the scale factor
$a$, the sum of the energy density and pressure $(\rho+p)$ and the
energy density $\rho$ during the bounce period for logarithmic
corrected model in fractal universe. The parameters are chosen for
this case as $k=G=\alpha'=\beta'=\Lambda=1$ and $n=v_0=2$ and
$\gamma=0$ (blue line), $1/3$ (red line) and $-1/2$
(green line).\\

\vspace{4mm}

\end{figure}

\subsection{Power Law Corrected Model in Fractal Universe}

In this case, we get from \eqref{fld eqn 2 Fractal power law corr}

\begin{equation}
8\pi G(\rho_b+p_b)=\frac{k}{{a_b}^2}\left[2-\frac{\epsilon
k^{\frac{\epsilon}{2}-1}}{(a_b/r_c)^{\epsilon-2}}\right]-\frac{\ddot{a_b}}{a_b}\left[(n+2)-\frac{\epsilon
k^{\frac{\epsilon}{2}-1}}{(a_b/r_c)^{\epsilon-2}}\right]
\end{equation}

The right hand side of the above expression must be positive for
the validity of the NEC. Here the energy density during bounce is
given by

\begin{equation}
\rho_b=\frac{3}{8\pi
G}\left[\frac{k}{{a_b}^2}-\frac{\Lambda}{3}-\frac{k^{\frac{\epsilon}{2}}}{{a_b}^\epsilon{r_c}^{2-\epsilon}}\right]
\end{equation}

\subsubsection{Case I: $k=0$}

Here

\begin{equation}
\rho_b+p_b=-\frac{(n+2)}{8\pi G}\frac{\ddot{a_b}}{a_b}
\end{equation}

This means that NEC is satisfied when $n<-2$. In this case
$\rho_b=-\frac{\Lambda}{3\pi G}$ implying a solution without
favorable criteria for a bounce.\\

\subsubsection{Case II: $k=\pm1$}

The energy density in this case is given by

\begin{equation}
{\rho_b}_{|_{k=1}}=\frac{3}{8\pi
G}\left[\frac{1}{{a_b}^2}-\frac{\Lambda}{3}-\frac{1}{{a_b}^\epsilon{r_c}^{2-\epsilon}}\right]
\end{equation}

and

\begin{equation}
{\rho_b}_{|_{k=-1}}=-\frac{3}{8\pi
G}\left[\frac{1}{{a_b}^2}+\frac{\Lambda}{3}+\frac{(-1)^\frac{\epsilon}{2}}{{a_b}^\epsilon{r_c}^{2-\epsilon}}\right]
\end{equation}

It may be noted that the third term decides for a non negative
$\rho_b$. If $\epsilon=2m$, where $m$ is odd, then one can attain
a bouncing solution. Otherwise there will be no bounce.\\

The criteria for the validity of the NEC in the two cases
($k=\pm1$) are

\begin{equation}
2-\frac{\epsilon}{(a_b/r_c)^{\epsilon-2}}>{a_b}{\ddot{a_b}}\left[(n+2)-\frac{\epsilon}{(a_b/r_c)^{\epsilon-2}}\right],~~~k=1
\end{equation}

\begin{equation}
2+\frac{\epsilon(-1)^\frac{\epsilon}{2}}{(a_b/r_c)^{\epsilon-2}}+{a_b}{\ddot{a_b}}\left[(n+2)
+\frac{\epsilon(-1)^\frac{\epsilon}{2}}{(a_b/r_c)^{\epsilon-2}}\right]>0,~~~k=-1
\end{equation}

The evolution of the scale factor $a$, the sum of the energy
density and pressure $(\rho+p)$ and the energy density $\rho$
during the bounce period for power law corrected model in fractal
universe are shown in Figs. 14-16. A barotropic equation of state
$p=\gamma\rho$ is assumed for the matter content. The parameters
are chosen differently for generating the plots. For scale factor
we have chosen $k=G=\Lambda=1$, $n=2$, $v_0=0.2$, $r_c=10$,
$\epsilon=4$ and $\gamma=-1/2$ (both blue and red lines with
different initial conditions) and $\gamma=1$ (green line). Here
variations have been made both in terms of initial conditions and
choice of $\gamma$ to achieve a bouncing solution. However, the
plots for $\rho+p$ and $\rho$ were carried out with a different
choices of the parameter $\epsilon$, since it plays a crucial role
for the satisfaction of NEC. In these two cases, the initial
conditions are kept constant. The parameters are chosen for these
two cases as $k=-1$, $G=\Lambda=1$, $n=2$, $r_c=10$, $v_0=0.2$,
$\gamma=1/3$ and $\epsilon=6$ (green line), $10$ (red line) and
$14$ (blue line). it can be observed that the plots show the
satisfaction of the NEC.\\

\begin{figure}
\vspace{4mm}

\includegraphics[height=2.0in]{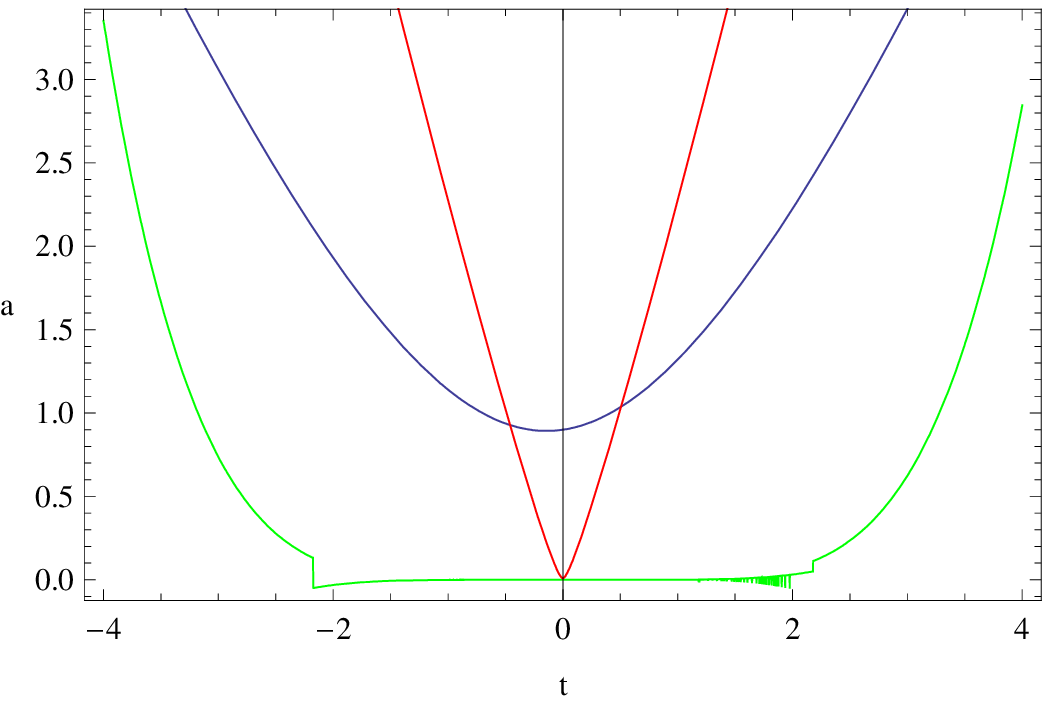}~~~~
\includegraphics[height=2.0in]{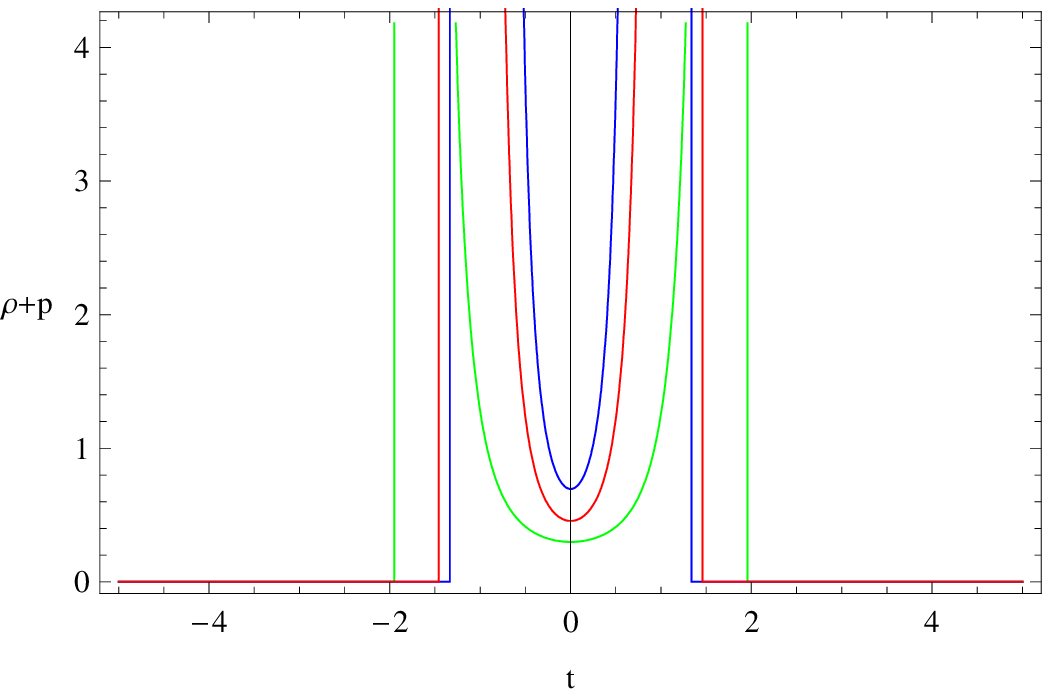}\\
\vspace{2mm} ~~~~~~~~~~~~~~~~~~~~~~~~~~~~~Fig.14~~~~~~~~~~~~~~~~~~~~~~~~~~~~~~~~~~~~~~~~~~~~~~~~~~~~~~~~~~~Fig.15~~~~\\
\vspace{4mm}

~~~~~~~~~~~~~~~~~~~~~~~~~~~~~~~~~~~~~\includegraphics[height=2.0in]{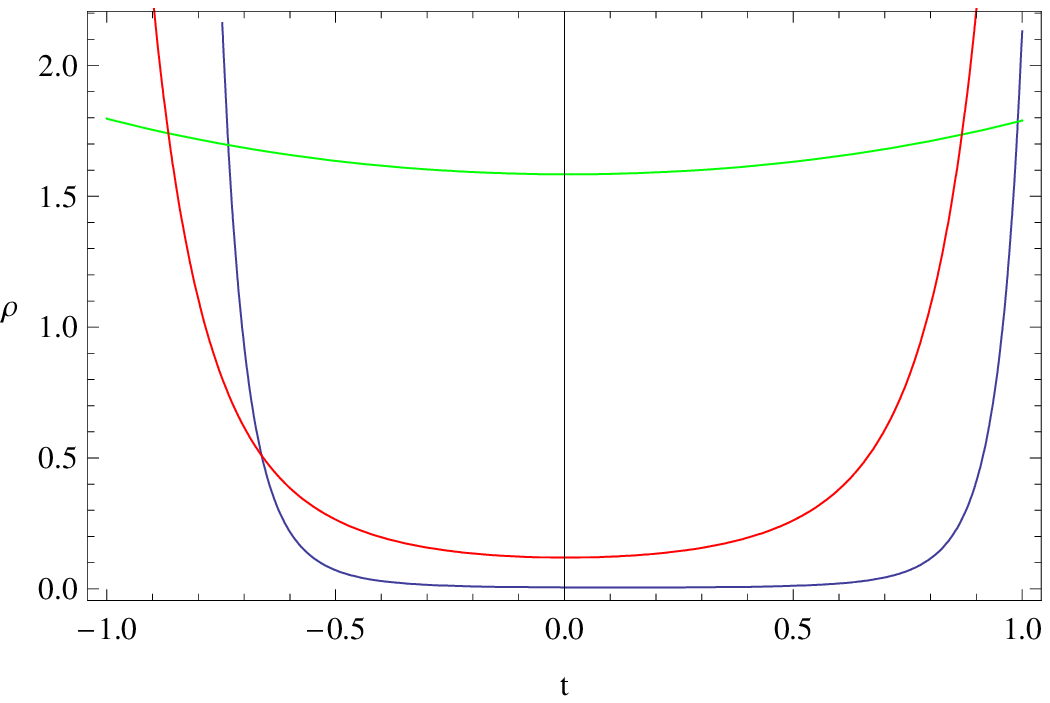}\\
\vspace{2mm} ~~~~~~~~~~~~~~~~~~~~~~~~~~~~~~~~~~~~~~~~~~~~~~~~~~~~~~~~~~~~~~~~~Fig.16~~~~~~~~~~~~~~\\

~~~Figs. 14-16 show the dynamical behaviors of the scale factor
$a$, the sum of the energy density and pressure $(\rho+p)$ and the
energy density $\rho$ during the bounce period for power law corrected model in fractal universe.\\

\vspace{4mm}

\end{figure}

\section{Discussions and Concluding Remarks}

In this work, the modified field equations in logarithmic and
power law versions of entropy corrected models in Einstein's
gravity in the background of FRW Universe are studied. After
discussing briefly about the Ho$\check{\text r}$ava-Lifshitz
gravity, we have formulated the modified field equations in
logarithmic and power law versions of entropy corrected models in
Ho$\check{\text r}$ava-Lifshitz gravity. We have performed the
stability analysis of the dynamical system for these models. For
logarithmic version of entropy corrected model in Ho$\check{\text
r}$ava-Lifshitz gravity, we have obtained a stable critical point
by assuming $k=-1$, $\Lambda=\alpha'=\beta'=1$, $\gamma=1/3$,
$\nu=1$. The graph of $a$ vs $H$ in this case is drawn in Fig.1
and we have found that the critical point is a stable center. For
power law version of entropy corrected model in Ho$\check{\text
r}$ava-Lifshitz gravity, we have obtained a stable critical point
by assuming $k=1$, $\Lambda=\nu=1$, $\epsilon=2$, $r_c=10$ and
$\gamma=1/3$. The phase space diagram of $a$ vs $H$
is shown in Fig.2 and a stable center is found.\\

In another section, we have briefly discussed about the fractal
universe and later have formulated the modified field equations in
logarithmic and power law versions of entropy corrected models in
fractal universe. The stability analysis for the dynamical system
is then performed for these models in the framework of fractal
universe. For logarithmic version of entropy corrected model in
fractal universe, we have found the stable critical point for the
choice of the parameters $k=1$, $\Lambda=\alpha'=\beta'=1$, $n=2$,
$v_0=0.2$, $\omega=0.3$ and $\gamma=0$. From the phase diagram
Fig. 3, we have got a stable center. For power law version of
entropy corrected model in fractal universe, a stable critical
point is obtained by choosing $k=-1$, $\Lambda=1$, $\epsilon=4$,
$r_c=10$, $n=2, v_{0}=0.5, \omega=0.6$ and $\gamma=-1/2$. The
phase space diagram of $a$ vs $H$ has been presented in Fig.4 and
a stable center is obtained.\\

Furthermore, we have analyzed the bouncing scenarios of the
universe in Ho$\check{\text r}$ava-Lifshitz gravity and fractal
model for both logarithmic and power law entropy corrected
versions in $k=0,+1,-1$ separately. For different cases, we have
examined the validity of the null energy condition (NEC) at the
time of bounce. For logarithmic corrected entropy in
Ho$\check{\text r}$ava-Lifshitz gravity with $k=0$, we have
observed that the NEC is violated. Also $\rho_b$ attained a
minimum value during bounce given by
$(\rho_{b})_{min}=-\frac{3\nu\Lambda}{16\pi{G_c}}$. So this case
has turned out to be unfavorable for bounce. On the other hand,
for $k=\pm 1$, the NEC may be satisfied with some conditions
provided in equations \eqref{cond at k=1} and \eqref{cond at
k=-1}. Also $\rho_b$ may attain maximum value during bounce, which
are given in equations \eqref{rho at k=1} and \eqref{rho at k=-1}.
In the case, the evolution of the scale factor $a$, $(\rho+p)$ and
$\rho$ during the bounce period are shown in Figs. 5-7. For viable
choices of the parameters involved, the plots shown an oscillating
universe with the NEC being satisfied.\\

For power law corrected entropy in Ho$\check{\text r}$ava-Lifshitz
gravity with $k=0$, we have observed that the NEC is violated and
at the time of bounce, the energy density becomes negative, which
shows there is no bounce. But for $k=\pm 1$ with $\epsilon=2m$,
where $m$ is odd, there would be a chance to have bounce. In all
other cases, we have unfavorable condition for bounce. In this
case, $\rho_{b}$ attains its maximum value. The bouncing solution,
the evolution of the scale factor $a$, $(\rho+p)$ and
$\rho$ during the bounce period are shown in Figs. 8-10.\\

For logarithmic corrected entropy in fractal universe with $k=0$,
we have observed that the NEC is satisfied for a negative $n$ and
violated otherwise. This case is not favorable for bounce as
$\rho_b$ becomes negative. But for $k \pm 1$, NEC may be satisfied
with some restrictions. The evolution of the scale factor $a$,
$(\rho+p)$ and $\rho$ during the bounce period are shown in Figs.
11-13. The plots clearly show a bouncing solution with the NEC
being satisfied.\\

For power law corrected entropy in fractal universe with $k=0$, we
have observed that the NEC is satisfied when $n<-2$. But this
solution indicates an unfavorable criteria for a bounce as
$\rho_b$ becomes negative. But for $k=\pm 1$ with $\epsilon=2m$,
where $m$ is odd, one can attain a bouncing solution. In this
case, $\rho_{b}$ attains its maximum value. The validity of NEC
depends on some restrictions. The evolution of the scale factor
$a$, $(\rho+p)$ and $\rho$ during the bounce period are shown in
Figs. 14-16. From the figures, it can be observed that the plots
show the satisfaction of the NEC.\\

{\bf Acknowledgement}: The author TB is thankful to IUCAA, Pune,
for their warm hospitality where part of the work has been done
during a visit under the Associateship Programme. \\\\

\end{document}